\documentclass[aps,twocolumn,pre,amsmath,floatfix,superscriptaddress]{revtex4-1}

\usepackage{graphicx}
\usepackage{times}

\usepackage[varg]{txfonts}
\usepackage{bm}
\usepackage{amsmath}
\usepackage{amsfonts}
\usepackage{amssymb}
\usepackage{xspace}
\usepackage{xr}
\usepackage{natbib} %\bibliographystyle{prsty}
\usepackage{xcolor}
\usepackage[caption=false]{subfig}

\usepackage[colorlinks,bookmarks=false,citecolor=blue,linkcolor=red,urlcolor=blue]{hyperref}

\usepackage{color}

\newcommand{\bra}[1]{\langle\left.{#1}\right|}
\newcommand{\ket}[1]{\left|{#1}\right.\rangle}
\newcommand{\beq}{\begin{equation}}
\newcommand{\eeq}{\end{equation}}

\newcommand{\la}{\left\langle}
\newcommand{\ra}{\right\rangle}
\newcommand{\lb}{\left[}
\newcommand{\rb}{\right]}
\newcommand{\lp}{\left(}
\newcommand{\rp}{\right)}

\newcommand{\be}{\begin{equation}}
\newcommand{\ee}{\end{equation}}
\renewcommand{\Re}{{\rm \, Re\,}}

\begin{document}

\title{Eigenstate Thermalization, Random Matrix Theory and Behemoths}

\author{Ivan M. Khaymovich}

\affiliation{Max Planck Institute for the Physics of Complex Systems, N\"{o}thnitzer Str. 38, 01187 Dresden}
\affiliation{Institute for Physics of Microstructures, Russian Academy of Sciences, 603950 Nizhny Novgorod, GSP-105, Russia}

\author{Masudul Haque}

\affiliation{Max Planck Institute for the Physics of Complex Systems, N\"{o}thnitzer Str. 38, 01187 Dresden}
\affiliation{Department of Theoretical Physics, Maynooth University, Co.\ Kildare, Ireland.}

\author{Paul A. McClarty}

\affiliation{Max Planck Institute for the Physics of Complex Systems, N\"{o}thnitzer Str. 38, 01187 Dresden}

\pacs{}

\begin{abstract}

The eigenstate thermalization hypothesis (ETH) is one of the cornerstones in our understanding of
quantum statistical mechanics. The extent to which ETH holds for nonlocal operators is an open
question that we partially address in this paper. We report on the construction of highly nonlocal
operators, Behemoths, that are building blocks for various kinds of local and non-local
operators. The Behemoths have a singular distribution and width $w\sim \mathcal{D}^{-1}$
($\mathcal{D}$ being the Hilbert space dimension).  From them, one may construct local operators
with the ordinary Gaussian distribution and $w\sim \mathcal{D}^{-1/2}$ in agreement with ETH.
Extrapolation to even larger widths predicts sub-ETH behavior of typical nonlocal operators with
$w\sim \mathcal{D}^{-\delta}$, $0<\delta<1/2$. This operator construction is based on a deep analogy
with random matrix theory and shows striking agreement with numerical simulations of non-integrable
many-body systems.

\end{abstract}

\maketitle

%\date{}      

%%%%%%%%%%%%%%%%%%%%%%%%%%%%%%%%%%%%%%%%%%%%%%%%%%%%%%%%%%%%%%%%%%
%%%% INTRODUCTION
%%%%%%%%%%%%%%%%%%%%%%%%%%%%%%%%%%%%%%%%%%%%%%%%%%%%%%%%%%%%%%%%%%

{\it Introduction} -- Some of the most fundamental questions in quantum statistical mechanics relate
to whether and how thermalization occurs in isolated quantum systems out of equilibrium. Whereas a
closed quantum system in a pure state never comes to thermal equilibrium, subsystems may thermalize
in the sense that observables acting on the subsystem may be computed from a thermal ensemble in the
long time limit.  The process of thermalization depends on the nature of the many-body system, the
initial state, the subsystem and the observable.  Despite the complexity of this problem, the
eigenstate thermalization hypothesis (ETH) boils the issue down to the nature of the matrix element
distribution of the observable in the eigenstate basis.
ETH is the conjecture that the fluctuations of these matrix elements are exponentially small in the
system size
\cite{berry1977regular,pechukas1983distribution,pechukas1984remarks,peres1984ergodicity, feingold1984ergodicity,feingold1986distribution,jensen1985statistical,PhysRevE.50.888,PhysRevA.43.2046,d2016quantum,
  NandkishoreHuse_AnnuRev2015}.
Denoting eigenvalues and eigenstates by $\mathsf{E}_A$ and $\vert \mathsf{E}_A\rangle$, ETH for an
operator $\hat{O}$ is stated as
\begin{equation} 
\la \mathsf{E}_A\left| \hat{O}\right| \mathsf{E}_B \ra = \delta_{AB}
f^{(1)}_{O}(\bar{\mathsf{E}}) + e^{-S(\bar{\mathsf{E}})/2} f^{(2)}_{O}(\bar{\mathsf{E}},\omega)
R_{AB}
\label{eq:ETH}
\eeq
where $S\sim \log \mathcal{D}$ is the entropy and $\mathcal{D}$ is the Hilbert space dimension,
$\bar{\mathsf{E}}=(1/2)(\mathsf{E}_A + \mathsf{E}_B)$ and $\omega=\mathsf{E}_B - \mathsf{E}_A$,
$R_{AB}$ is a random variable with zero mean and unit variance, and $f^{(1,2)}$ are smooth
functions. If condition (\ref{eq:ETH}) holds then the long time average of $\hat{O}$ matches the
thermal result
\cite{berry1977regular,pechukas1983distribution,pechukas1984remarks,peres1984ergodicity,
  feingold1984ergodicity,feingold1986distribution, jensen1985statistical, PhysRevE.50.888,
  PhysRevA.43.2046, d2016quantum, NandkishoreHuse_AnnuRev2015}. A crucial aspect of
ETH is the scaling of the width of the operator distribution: the width of the distribution falls
off as $e^{-S(\bar{\mathsf{E}})/2}\sim \mathcal{D}^{-1/2}$.  This scaling is based on the similarity
between typical many-body eigenstates and random states \cite{Marquardt_PRE12,
  Beugeling_scaling_PRE14, Beugeling_offdiag_PRE2015}.

\begin{figure}[]
\centering
%\hspace{-.5cm}
\centering
\includegraphics[width=0.95\columnwidth]{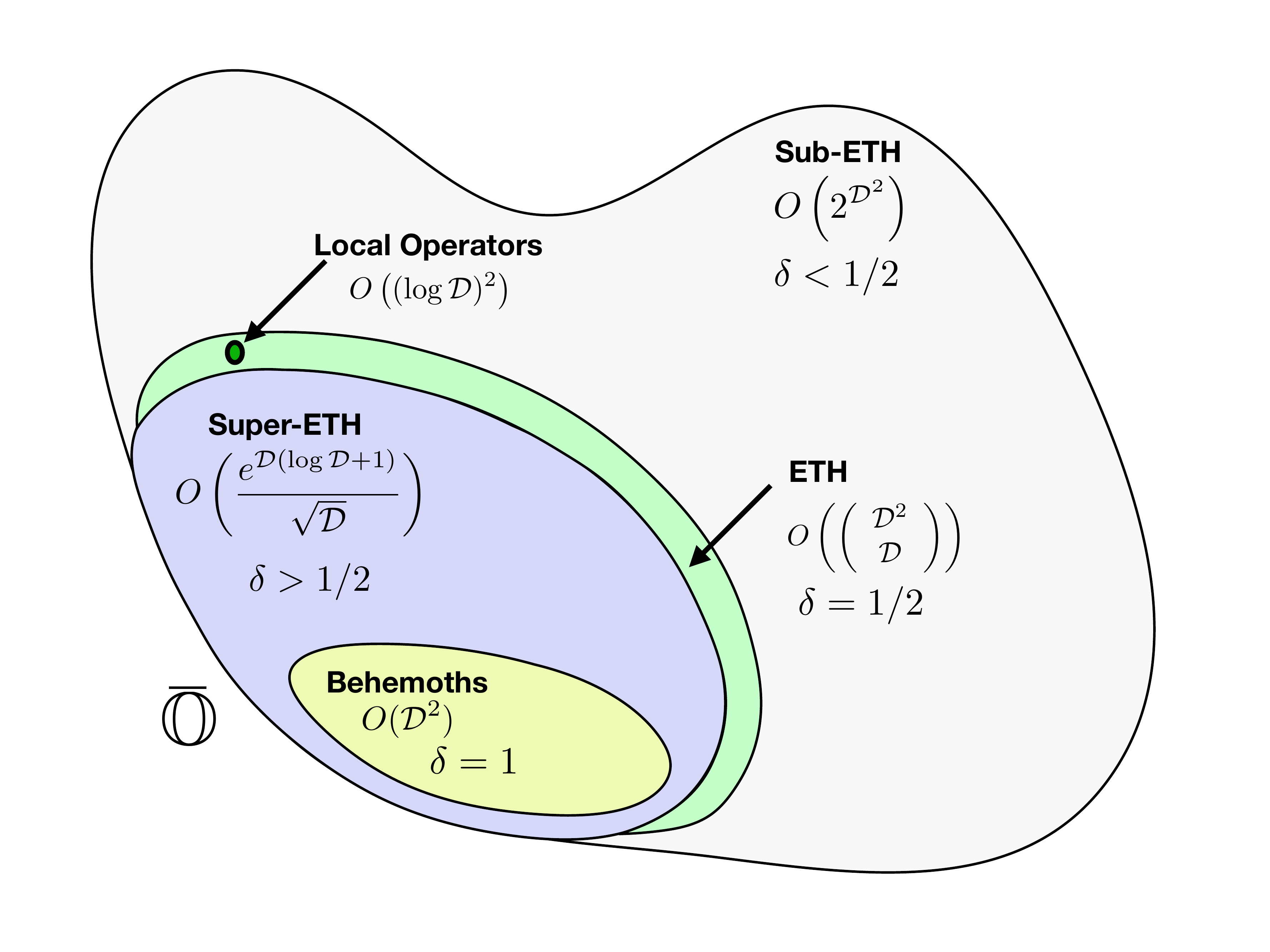}
\caption{Schematic showing the space of operators $\bar{\mathbb{O}}$ having elements zero or unity in the
  configuration basis.  The operators are organized into classes distinguished by the scaling of the
  width $\sigma\sim \mathcal{D}^{-\delta}$ of matrix element distributions in the eigenstate basis.
\label{fig:1}
}
\end{figure}

The weight of evidence based on a large number of numerical studies strongly suggests that ETH is
satisfied for typical states of generic nonintegrable systems and for physical observables
\cite{d2016quantum, Rigol_Nature2008, Rigol_PRL2009, BiroliKollathLauchli_PRL10, RigolSantos_PRA10,
  SantosRigol_PRE10, Roux2010PRA81, Motohashi2011, Marquardt_PRE12, BrandinoKonikMussardo_PRB12,
  IkedaUeda_PRE13, SteinigewegPrelovsek_PRE13, IkedaHuse_PRL14, Beugeling_scaling_PRE14,
  SorgVidmarHeidrichMeisner_PRA14, SteinigewegGogolinGemmer_PRL2014, Beugeling_offdiag_PRE2015, FratusSrednicki_PRE2015, 
  MondainiSrednickiRigol_PRE16, ChandranBurnell_PRB16, LuitzBarlev_PRL16,
  NandkishoreHuse_AnnuRev2015, MondainiRigol_PRE2017, 2017arXiv171208790M}.
However, there is currently little sharp understanding of the class of operators which satisfy ETH.
While local observables are expected to obey ETH, one might imagine that sufficiently nonlocal
operators are athermal because there is no distinction between the subsystem and the bath.
Projection operators onto eigenstates are extreme examples of this type since these take values $0$
or $1$ while the thermal expectation value of such an operator is exponentially small in the system
size. Earlier work that has touched on this question includes Refs.~\cite{PhysRevLett.120.080603,
  garrison2018, ChandranBurnell_PRB16,goldstein2014strong, Beugeling_entanglement_JSM2015}.

In this paper, we explore a correspondence between random matrix theory (RMT) and many-particle
quantum systems that allows one to make testable predictions for the scaling of matrix element
distributions of fairly general operators. Fig.~\ref{fig:1} summarizes our classification of
operators. We begin by considering a class of highly nonlocal operators that connect single pairs of
many-body configurations.  We will call these Behemoth operators.
Using RMT, we derive analytical predictions for the distribution of eigenstate matrix elements of
Behemoths.  We demonstrate that Behemoths in a wide class of lattice many-body systems match the RMT
predictions.  We show that these operators are distinguished by exhibiting {\it super}-ETH scaling
with eigenstate distribution width scaling as $\mathcal{D}^{-1}$.

The Behemoth operators have a deeper importance: they are building blocks for a vastly larger class
of operators that includes the local operators. By connecting more and more pairs of many-body
configurations, one can tune the scaling of the matrix element distribution to be
$\mathcal{D}^{-\delta}$. The {\it super}-ETH operators have $1/2<\delta\leq 1$, the operators that
obey ETH (such as local operators) have $\delta =1/2$ and the {\it sub}-ETH operators have
$\delta<1/2$.  As with Behemoth operators, RMT supplies predictions for the distribution of all such
operators that we compare with numerical results for many-body Hamiltonians.  This construction
is an alternative route to the $\mathcal{D}^{-1/2}$ (ETH) scaling of local
operators.

%%%%%%%%%%%%%%%%%%%%%%%%%%%%%%%%%%%%%%%%%%%%%%%%%%%%%%%%%%%%%%%%%%
%%%% ANALOGY
%%%%%%%%%%%%%%%%%%%%%%%%%%%%%%%%%%%%%%%%%%%%%%%%%%%%%%%%%%%%%%%%%%

{\it Analogy between Random Matrix Theory and Many-Body Physics} -- Suppose $H_{ij}$ is a $N\times
N$ random matrix with eigenstates $\ket{E_{\alpha}}$.  We interpret $H_{ij}$ as a fully-connected
single particle hopping Hamiltonian.  Then $i,j$ are `site' indices.  Also,
$\mathsf{H}_{\boldsymbol{n} \boldsymbol{n'}}$ is a $\mathcal{D}\times\mathcal{D}$ many-body lattice
Hamiltonian with eigenstates $\vert \mathsf{E}_{A}\rangle$.  Each basis state $\boldsymbol{n}$ is a
many-body configuration, specified by the occupancies of the $L$ sites in the lattice.

For the RMT, we consider $\hat{\omega}_{ij}\equiv \hat{d}_i^\dagger \hat{d}_j$, the single particle
hopping operators between sites $j$ and $i$. 
In the many-body model, the analogous operators connect pairs of configurations
$\boldsymbol{n}$ and $\boldsymbol{n}'$ in the occupation number basis: 
\begin{equation}
\label{eq:MBS_Behemoth_defn}  
\hat{\Omega}_{\bf n n'}
\equiv \vert \boldsymbol{n}\rangle\langle \boldsymbol{n}'\vert.
\end{equation}
As these are extremely nonlocal, we call them Behemoth operators.
In the configuration basis, the matrices representing Behemoth operators have a single nonzero
entry.  Behemoths thus form a natural basis for all operators.
Hermitian Behemoths are defined as $\hat{\Gamma}_{\bf n n'} \equiv \hat{\Omega}_{\bf n n'}
+\hat{\Omega}_{\bf n' n} $.

We will examine the distribution of eigenstate matrix elements of Behemoths.  We propose that the
statistics of such many-body matrix elements match those of the matrix elements of
$\hat{\omega}_{ij}= \hat{d}_i^\dagger \hat{d}_j$ in RMT.
Below, we calculate their distribution on the RMT side and then carry out numerical tests of
the correspondence.

If the many-body Hamiltonian conserves particle number, $N_{\rm p}$, then
for spinless fermions or hard-core bosons the many-body matrix elements of the
Behemoths are 
\begin{equation}
\label{eq:MBS_Xab_nn'}
\la \mathsf{E}_A \left|
 \hat{\Omega}_{\bf n n'}  \right| \mathsf{E}_B\ra \equiv \la \mathsf{E}_A \left|
\prod_{k=1}^{m} \hat c_{i_k}^\dagger \hat c_{j_k}
\prod_{l=1}^{N_{\rm p}-m}\hat c_{p_l}^\dagger \hat c_{p_l}
\right| \mathsf{E}_B\ra  \ .
\end{equation}
The Behemoth changes one configuration of $N_{\rm p}$ particles into another, by moving $m$
particles from one set of sites to another.  ($\{j_k\}$ are occupied sites in the $\boldsymbol{n}$
configuration and empty sites in the $\boldsymbol{n'}$ configuration, and vice versa for the
$\{i_k\}$ sites.)  The other $N_{\rm p}-m$ particles do not need to be moved; $\{p_k\}$ are occupied
sites in both configurations.  For spin-1/2 systems, spins up/down are interpreted as occupied/empty
sites and $N_{\rm p}$ is the number of up spins.  Eq.\ \eqref{eq:MBS_Xab_nn'} can be readily
generalized to cases where multiple occupancies are allowed (e.g., bosonic or fermionic Hubbard
models, or $S>\frac{1}{2}$ spin systems), and to systems where particle number is not conserved.

{\it From Nonlocal to Local} --
Besides $\hat{\Omega}_{\bf n n'}$, we consider operators with varying degrees of locality,
$\hat{\Omega}_{M} = \prod_{k=1}^n \hat c_{i_k}^\dagger \hat c_{j_k}$, which hop $n$ of the $N_{\rm
  p}$ particles ($n\lesssim N_{\rm p}$).
The expectation values of $\hat{\Omega}_{M}$ are $(2n)$-point correlators.  (For simplicity we
consider the sets $\{i_k\}$ and $\{j_k\}$ to have no intersection.)  Whereas $\hat{\Omega}_{\bf n
  n'}$ couples exactly two configurations, $\hat{\Omega}_{M}$ changes the configuration on $2n$
sites while the remaining sites may adopt any of $M\equiv \left( \begin{array}{c} L - 2n \\ N_{\rm
    p} - n \end{array} \right)$ configurations.  The matrix representing $\hat{\Omega}_{M}$ thus has
$M$ nonzero elements, each equal to $1$, i.e., $\hat{\Omega}_{M}$ is a sum of $M$ Behemoths.
The Behemoths correspond to $n=N_{\rm p}$, with $M=1(2)$ for non-hermitian (hermitian) cases.
The limit of a local single particle hopping operator is $n=1$.  Local operators are thus formed by
combining $M=O(\mathcal{D})$ Behemoths.

%%%%%%%%%%%%%%%%%%%%%%%%%%%%%%%%%%%%%%%%%%%%%%%%%%%%%%%%%%%%%%%%%%
%%%% RMT PREDICTIONS
%%%%%%%%%%%%%%%%%%%%%%%%%%%%%%%%%%%%%%%%%%%%%%%%%%%%%%%%%%%%%%%%%%

{\it Statistics of Many-Body Operators from RMT} -- We now make concrete predictions using RMT. The
RMT objects corresponding to the matrix elements of Eq.~(\ref{eq:MBS_Xab_nn'}) are $
\omega_{ij}^{\alpha\beta} = \la E_{\alpha} \left| \hat{d}_i^\dagger \hat{d}_j \right| E_{\beta}\ra =
u^{\star}_{\alpha,i}u_{\beta,j}$, where $u_{\alpha,i}\equiv \la i \vert E_{\alpha}\ra$.

We first concentrate on Gaussian orthogonal ensemble (GOE) matrices.  For sufficiently large matrix
sizes $N$, coefficients of eigenstates $u_{n,i}$ are real-valued independent Gaussian variables with
zero mean and variance $\sigma_1^2 = 1/N$ \cite{mehta2004random, beenakker1997random,
  guhr1998random, evers2008anderson},
The distribution is $P_u(u) = e^{-u^2/2\sigma_1^2}/\sqrt{2\pi \sigma_1^2}$.
Within this approximation, both the diagonal and off-diagonal matrix elements of
$\hat{\omega}_{ij}^{\alpha\beta}$ have the distribution
\begin{equation}
P_\omega(x) = \int_{-\infty}^\infty du_1 du_2 P_u(u_1) P_u(u_2) \delta(x - u_1 u_2)  = \frac{1}{\pi \sigma_1^2} K_0\lp \frac{|x|}{\sigma_1^2}\rp.
\label{eq:P_x_GOE}
\end{equation}
Here $K_\nu(x)$ is the modified Bessel function of the second
kind. For the RMT analogue 
$\gamma_{ij}^{\alpha\beta} = \la E_{\alpha} \left| \hat{d}_i^\dagger \hat{d}_j + \hat{d}_j^\dagger
\hat{d}_i \right| E_{\beta}\ra $ of the hermitian operator $\hat{\Gamma}^{AB}_{\bf n n'}$ we
distinguish between diagonal matrix elements ($\alpha=\beta$) for which we obtain $P_{\gamma, {\rm
    diag}}(y)=P_\omega(y/2)/2$ and off-diagonal matrix elements ($\alpha\neq\beta$) for which we must convolve
two distributions of the form (\ref{eq:P_x_GOE}) giving \beq P_\gamma(y) =
\frac{1}{2\pi}\int_{-\infty}^\infty e^{- i\omega y} \frac{d\omega}{1+\sigma_1^4\omega^2} =
\frac{e^{-|y|/\sigma_1^2}}{2\sigma_1^2}.
\label{eq:P_x_GOE_Hermitian}
\eeq

Next we look at sums of $M$ operators of type $\hat{\omega}_{ij}$ and calculate the distribution of
diagonal and off-diagonal matrix elements. The distribution of the sum $\omega_M \equiv \sum^M_k
\omega_{k} $ may be obtained from the Fourier transform
$\tilde{P}_{\omega_M}(q)=\int_{-\infty}^{\infty} e^{iq X} P_{\omega_M} (X) dX$ by taking the $M$-th
power of the $\tilde{P}_{\omega_M}$ distribution \cite{SM}, leading to
\begin{equation} P_{\omega_M}(X) = \frac{1}{\sqrt{\pi}\Gamma[M/2]
  \sigma_1^{2}}\lp\frac{|X|}{2\sigma_1^2}\rp^{\frac{M-1}2} K_{\frac{1-M}2}\lp
\frac{|X|}{\sigma_1^2}\rp  . 
\label{eq:P_x_GOE_M}
\end{equation}
This function is Gaussian for large enough $M$: $P_{\omega_M} (X) \approx
e^{-X^2/(2M\sigma_1^4)}/\sqrt{2\pi M}\sigma_1^2$, in accordance with the central limit theorem. The
variance of this distribution is $M\sigma_1^2 \sim MN^{-2}$ which goes as $1/N$ for $M \sim N$. 

The distribution of the hermitian analog, $\hat{\gamma}_{M'}$ for off-diagonal matrix elements
is Eq.~(\ref{eq:P_x_GOE_M}) with $M=2M'$. The distribution for diagonal elements of
$\hat{\gamma}_{M'}$ is $P_{\omega_M}(Y/2) $ with $M=M'$.

The analysis for the GUE case is similar \cite{SM}.  The off-diagonal
matrix elements are now complex; the marginal distributions for real and imaginary parts of
$\omega_{ij}^{\alpha\beta}$ have exponential form.  The amplitude has the distribution
\begin{equation}
P_{\vert\omega\vert}(x) = \frac{x }{\sigma_2^4} K_0 \left( \frac{x}{\sigma_2^2}  \right)
\label{eq:P_x_GUE}
\end{equation}
which vanishes for $x\to0$.  Here $\sigma_2^2=1/(2N)$.  Other GUE and GSE distributions are derived
for completeness in \cite{SM}.

We now discuss these results in the light of the above-mentioned correspondence with many-body
physics.  For eigenstates in the middle of the spectrum of a local nonintegrable model - those for
which the energy dependence of the states is weakest - we expect that the off-diagonal matrix
elements of Behemoth operators of the type \eqref{eq:MBS_Xab_nn'}) should be distributed according
to \eqref{eq:P_x_GOE}, or according to \eqref{eq:P_x_GUE} if time reversal symmetry is violated.
Similarly, hermitian Behemoths and diagonal matrix elements should follow the RMT distributions
outlined above. 
The width $\sigma_1^2 = 1/N$ in RMT becomes $1/\mathcal{D}$ in the many-body case.  The Behemoths
thus obey a {\it super-ETH} scaling behavior.  Then, by tuning $M$ in Eq.~(\ref{eq:P_x_GOE_M}) we
interpolate between Behemoth operators for $M=1$ to local one-particle hopping operators for
$M=\left( \begin{array}{c} L - 2 \\ N_{\rm p} - 1 \end{array} \right)$ where there is particle
number conservation and $M=2^{L-2}$ otherwise. The width of local operators varies as
$\sqrt{M\mathcal{D}^{-2}}\sim \mathcal{D}^{-1/2}$ as enshrined in the usual statement of ETH.  Here,
we have made predictions for the whole distributions of classes of local and nonlocal operators with
no fitting parameters.
%

%%%%%%%%%%%%%%%%%%%%%%%%%%%%%%%%%%%%%%%%%%%%%%%%%%%%%%%%%%%%%%%%%%
%%%% NUMERICAL RESULTS
%%%%%%%%%%%%%%%%%%%%%%%%%%%%%%%%%%%%%%%%%%%%%%%%%%%%%%%%%%%%%%%%%%

%%%%%%%%%%%%%%%%%%%%%%%  FIGURE BEGINS %%%%%%%%%%%%%%%%%%%%%%%%%%%%%%%%%%%%%%%%%%
\begin{figure}[]
\centering
\hspace{-.5cm}
\centering
\includegraphics[width=0.95\columnwidth]{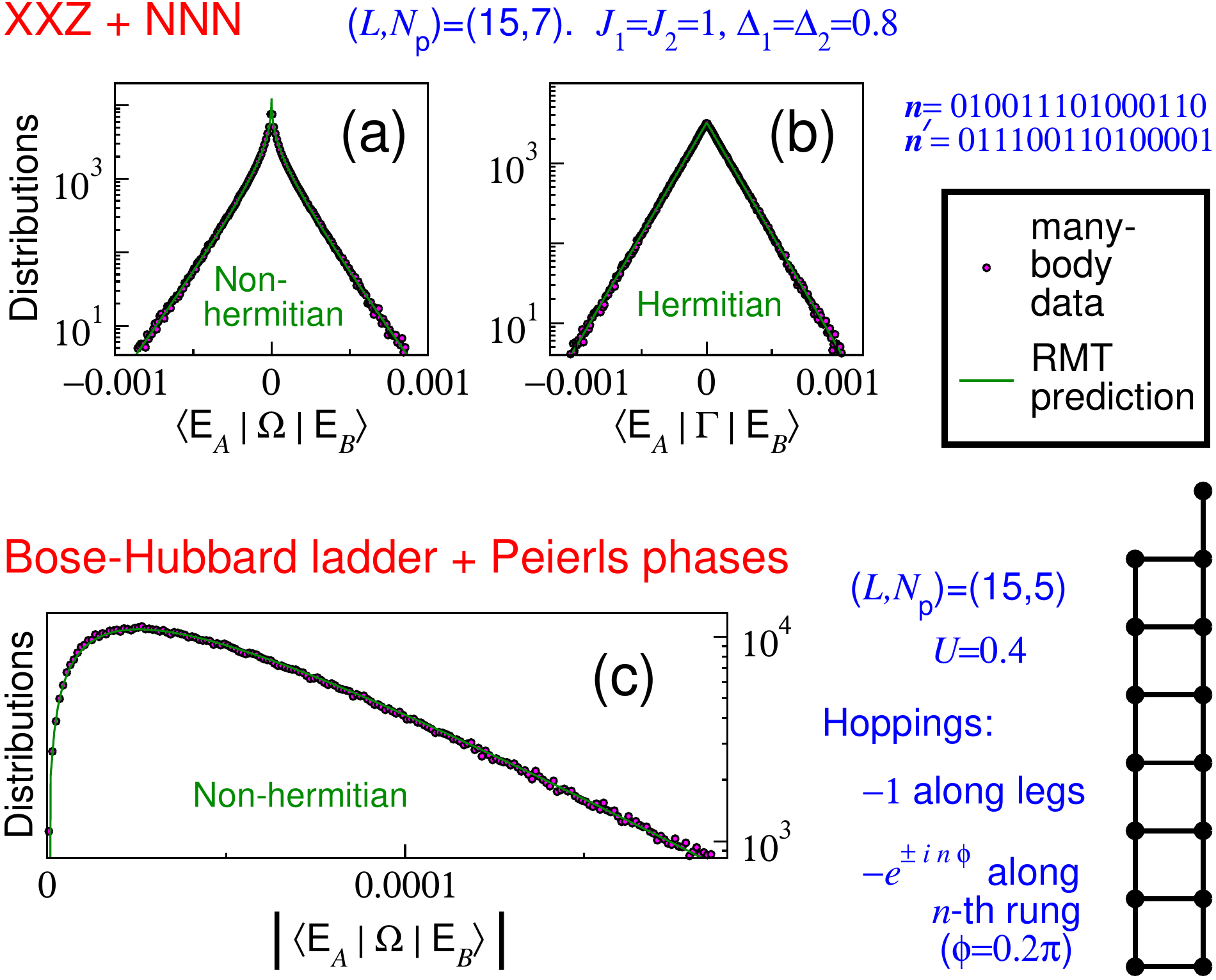}
\caption{Probability distributions of matrix elements of Behemoth operators, for two different
  many-body systems, compared with GOE and GUE predictions.  (a,b) Spin-1/2 chain with anisotropic
  Heisenberg ($XXZ$) couplings. Nearest-neighbor (NN) and next-nearest-neighbor (NNN) coupling
  strengths ($J_{1,2}$) and anisotropies ($\Delta_{1,2}$) are indicated.  (c) Bose-Hubbard ladder
  (geometry in sketch) subject to magnetic field.  Solid lines in (a,b,c) are predictions
  from Eqs.\ \eqref{eq:P_x_GOE}, \eqref{eq:P_x_GOE_Hermitian}, \eqref{eq:P_x_GUE} respectively.
\label{fig:2}
}
\end{figure}
%%%%%%%%%%%%%%%%%%%%%%%  FIGURE ENDS %%%%%%%%%%%%%%%%%%%%%%%%%%%%%%%%%%%%%%%%%%

{\it Numerical Results} -- We now present numerical tests of the conjectures described above.  We
performed these tests on an array of different interacting many-body lattice systems, including
spin-1/2 chains, bosonic Hubbard models and interacting spinless fermions.  Data for three different
systems appear in Figs.\ \ref{fig:2} and \ref{fig:4} while further comparisons (with specifications
of the models) appear in \cite{SM}.
Fig.~\ref{fig:2} shows the computed distributions (histograms) of off-diagonal matrix elements of
Behemoth operators for a GOE case (spin chain) and a GUE case (Bose-Hubbard ladder with a magnetic
field piercing every plaquette).  Fig.~\ref{fig:2}(a,b) uses a single Behemoth and 20\% of the
mid-spectrum eigenstates of the system.  Because particular operators may have atypical behavior, in
Fig.~\ref{fig:2}(c) and the rest of the paper we use statistics from a random collection of between
$50$ and $500$ Behemoths, the matrix elements are typically calculated between the central $50-200$
eigenstates.
Owing to the greater abundance of data for off-diagonal matrix elements we present these here and
show results for diagonal matrix elements - which have the same scaling - in \cite{SM}.

The agreement in Fig.~\ref{fig:2} with RMT predictions, Eqs.~(\ref{eq:P_x_GOE}),
(\ref{eq:P_x_GOE_Hermitian}), (\ref{eq:P_x_GUE}), is excellent.  The same is true for all systems we
have tested, for both off-diagonal and diagonal matrix elements \cite{SM}, as long as the
Hamiltonian parameters are in non-integrable (ergodic) regimes.

%%%%%%%%%%%%%%%%%%%%%%%  FIGURE BEGINS %%%%%%%%%%%%%%%%%%%%%%%%%%%%%%%%%%%%%%%%%%
\begin{figure}[]
\centering
\hspace{-.5cm}
\centering
\includegraphics[width=0.95\columnwidth]{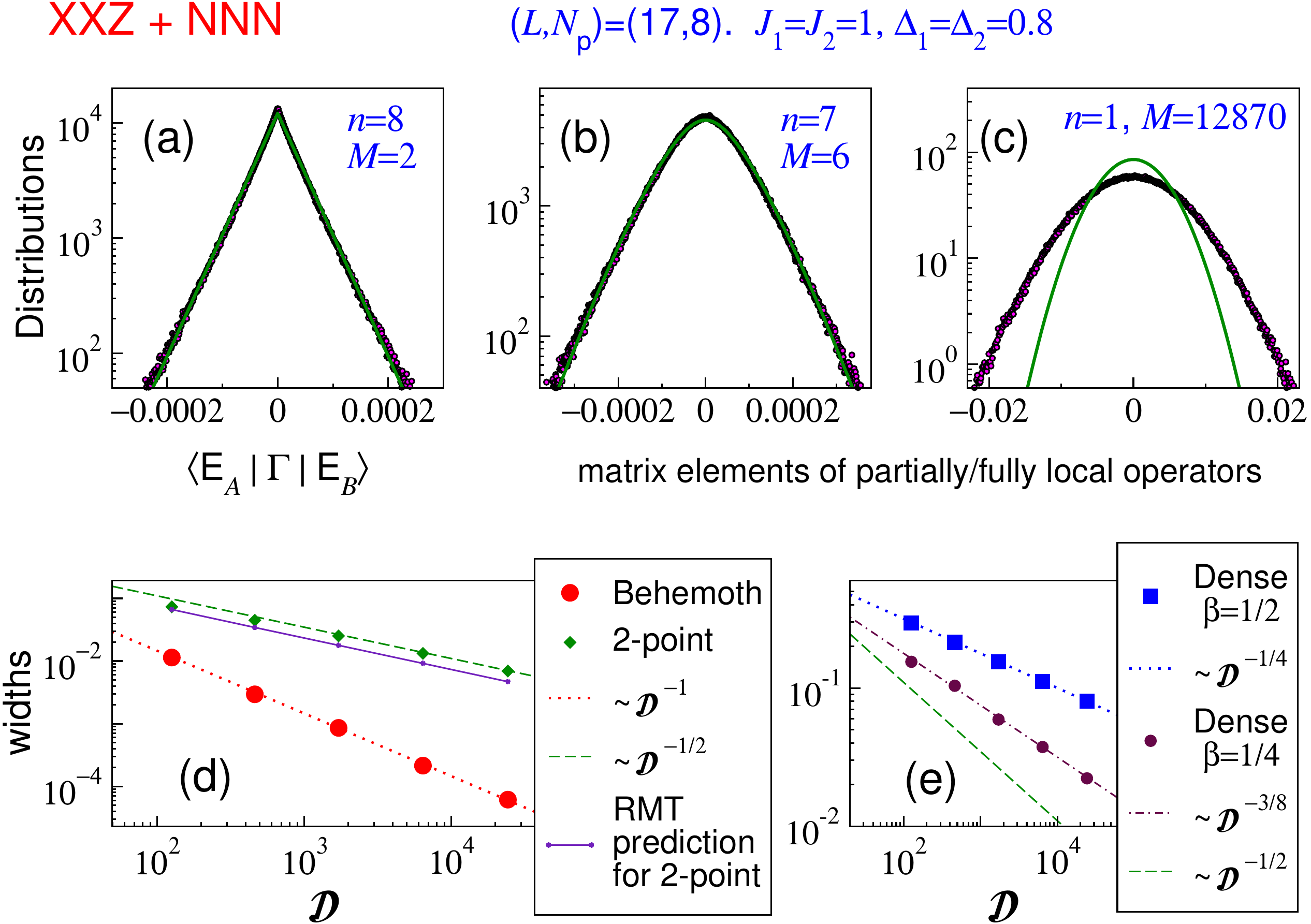}
\caption{(a,b,c) Distributions for Hermitian operators ($2n$-point correlators) of varying locality,
  from Behemoth (a) to 2-point correlator (c). 
  Number of terms $M$ in the operator matrix are shown.
Solid lines are RMT predictions, Eq.~(\ref{eq:P_x_GOE_M}).  (d) Width of distributions for Behemoths
and local operators.  The $\sim\mathcal{D}^{-1}$ line is the RMT prediction for Behemoths,
Eq.~\eqref{eq:P_x_GOE_Hermitian}.  RMT prediction for local operators (solid line) falls below the
data, consistent with panel (c).  (e) Width of distributions for two dense operators with
$M=\mathcal{D}^{1+\beta}$, showing the predicted sub-ETH scaling.  A dashed line for ETH scaling is
also shown.
\label{fig:3}
}
\end{figure}
%%%%%%%%%%%%%%%%%%%%%%%  FIGURE ENDS %%%%%%%%%%%%%%%%%%%%%%%%%%%%%%%%%%%%%%%%%%

We next consider operators interpolating between Behemoths and local operators, i.e., ($2n$)-point
correlators, with $n=N_{\rm p}$ for Behemoths and $n=1$ for local operators.  These correspond to
increasing $M$, the number of nonzero elements in the operator matrix.  Distributions of matrix
elements are shown in Fig.~\ref{fig:3}(a,b,c) for the spin chain, for $n=N_p$, $n=N_p-1$ and $n=1$.
The distribution goes from exponential to Gaussian as $M$ increases.  The scaling is
$\sim\mathcal{D}^{-1}$ (super-ETH) for Behemoths and $\sim\mathcal{D}^{-1/2}$ for $n=1$,
Fig.~\ref{fig:3}(d).

At moderate $M$ the agreement with Eq.~\eqref{eq:P_x_GOE_Hermitian} is excellent.  A striking effect
is seen at large $M$: the local operator distribution has the Gaussian shape and
$\mathcal{D}^{-1/2}$ scaling predicted by RMT, Eq.~\eqref{eq:P_x_GOE_Hermitian}, but the width is
systematically larger by a factor of order one (Fig.~\ref{fig:3}(c,d)).  This discrepancy is due to
the presence of weak correlations in the eigenstates \cite{SM}.  Correlation effects result in a
remarkable \emph{partial} violation of the central limit theorem.

Inverting the idea that $M<O(\mathcal{D})$ operators have super-ETH scaling, we now construct
operators with sub-ETH scaling.  By filling $M\sim \mathcal{D}^{1+\beta}$ elements ($\beta\in(0,1)$)
of the operator matrix, we obtain `dense' operators with matrix element distributions having widths
$\sim\sqrt{M}\mathcal{D}^{-1}\sim\mathcal{D}^{-1/2+\beta/2}$.  Two examples are shown in
Fig.~\ref{fig:3}(e); the predictions are borne out by the numerical results.

{\it Exceptions to RMT Scaling} -- 
We have shown that the correspondence between RMT and many-body operator distributions works very
well for the vast majority of eigenstates and typical Behemoths in nonintegrable models.  Under
exceptional circumstances, it can be made to fail. 
For example, if one or both of the configurations $\ket{\boldsymbol{n}}$, $\ket{\boldsymbol{n'}}$
are chosen such that they predominantly have weight in the highest-energy or lowest-energy
eigenstates, then the corresponding Behemoth $\Omega_{\boldsymbol{nn'}}$ will have anomalously small
matrix elements for eigenstates in the middle of the spectrum.  Maximally ferromagnetic
configurations for a spin chain can be used to construct such anomalies \cite{SM}.

The RMT correspondence is expected not to work in non-ergodic (ETH-violating) systems, e.g.,
many-body-localized (MBL) systems
\cite{PhysRevB.75.155111, NandkishoreHuse_AnnuRev2015, 2017arXiv171103145A,
  0295-5075-101-3-37003, PhysRevB.93.134201}
and integrable systems.  Fig.~\ref{fig:4}(a,b) shows the hermitian Behemoth distribution for an
interacting disordered system.  At small disorder (ergodic phase), the RMT-predicted exponential is
an excellent fit.  In the MBL phase, \ref{fig:4}(b), the distribution is a clear power law.  This
result immediately follows from the power law distribution of eigenstate coefficients known for the
MBL phase \cite{0295-5075-101-3-37003}.

In integrable systems, local operators have
non-ETH scaling (power-law with
system size) \cite{ziraldo2013relaxation, Beugeling_scaling_PRE14,
  Beugeling_offdiag_PRE2015,Alba_PRB15,ArnabSenArnabDas_PRB16, HaqueMcClarty_SYKETH}.  The
Behemoths, however, have the same $\mathcal{D}^{-1}$ scaling as in non-integrable cases, by
normalization.
Fig.~\ref{fig:4}(c) shows some deviation from the RMT prediction in the integrable $XXZ$ chain.  It
is conjectured that the coefficient distribution of integrable systems
approach a power law for $\mathcal{D}\to\infty$
\cite{beugeling2017statistical}, which implies that the Behemoth
distribution also approaches power law behavior.  The size-dependence of our data is consistent with this conjecture.

%%%%%%%%%%%%%%%%%%%%%%%  FIGURE BEGINS %%%%%%%%%%%%%%%%%%%%%%%%%%%%%%%%%%%%%%%%%%
\begin{figure}[]
\centering
\hspace{-.5cm}
\centering
\includegraphics[width=0.95\columnwidth]{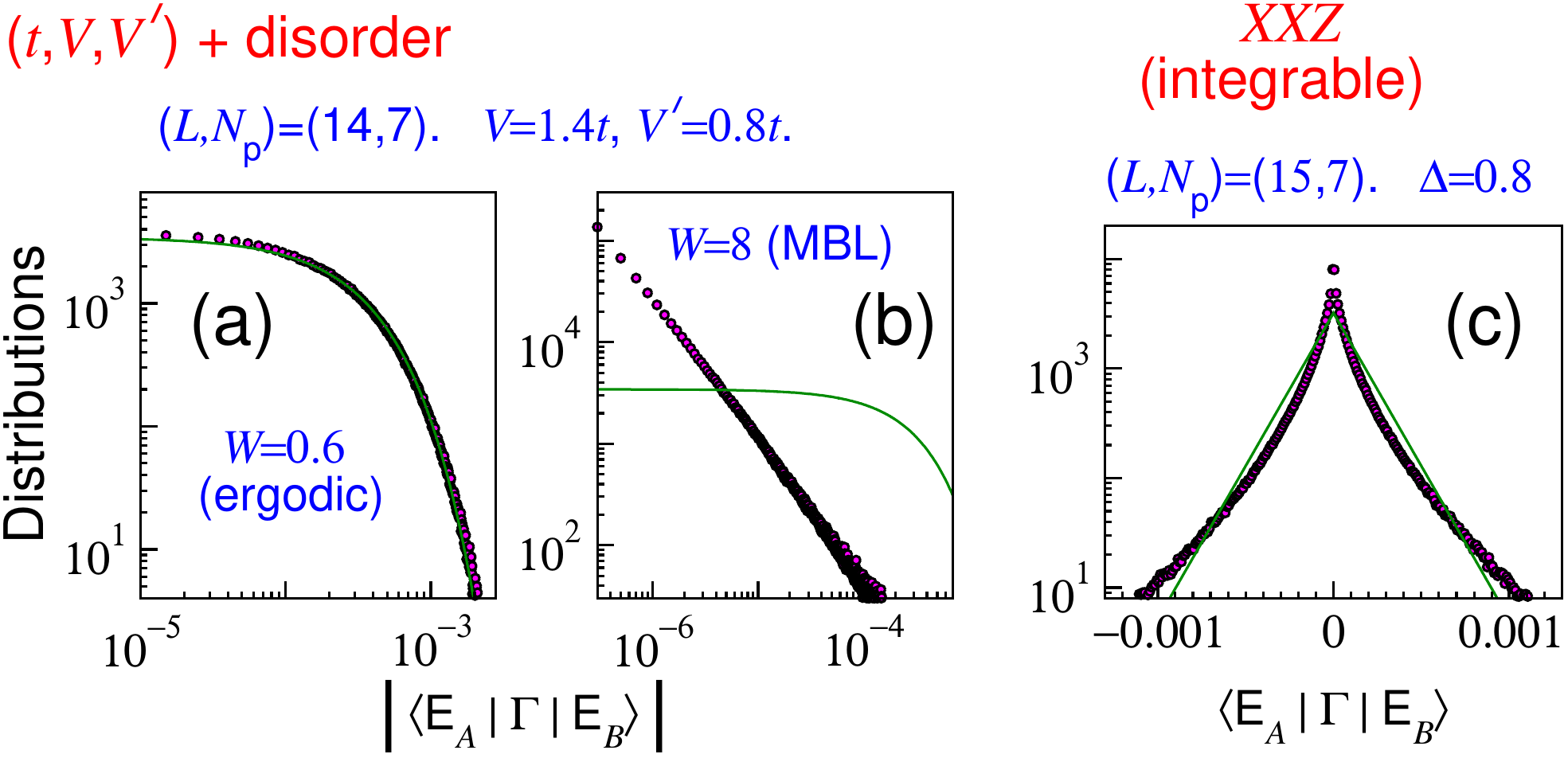}
\caption{\label{fig:4} Distributions for hermitian Behemoths.  (a,b) A spinless-fermion chain with
  NN and NNN interactions $V$ and $V'$, subject to Gaussian disorder of strength $W$.  (a) In the
  ergodic phase, distribution is exponential as predicted by RMT, Eq.~(\ref{eq:P_x_GOE_Hermitian}).
  (b) In the many-body localized phase, the distribution is a power law. (c) Integrable XXZ spin
  chain, showing deviation from RMT prediction.  }
\end{figure}
%%%%%%%%%%%%%%%%%%%%%%%  FIGURE ENDS %%%%%%%%%%%%%%%%%%%%%%%%%%%%%%%%%%%%%%%%%%

{\it Discussion} -- In this paper, we have investigated the matrix element distribution of operators
acting on typical (infinite temperature) eigenstates of many-body Hamiltonians. The distributions in
nonintegrable many-body interacting models largely match random matrix theory predictions.  We have
(i) constructed extremely nonlocal operators - Behemoths - that satisfy {\it super}-ETH scaling
(width $\sigma \sim \mathcal{D}^{-1}$ compared to $\sigma \sim \mathcal{D}^{-1/2}$ for ETH) , (ii)
interpolated between Behemoths and local operators noting that the form of the distribution and its
scaling can be captured by RMT but that for local operators there are small departures in the width
coming from correlations in the many-body states, (iii) obtained a set of typical operators with {\it
  sub}-ETH scaling ($\sigma \sim \mathcal{D}^{-\delta}$ with $\delta <1/2$).

In closing, we consider the frequency with which different scalings occur in the space of all
operators $\mathbb{O}$ acting on the many-body Hilbert space (Fig.~\ref{fig:1}). Consider a
many-body system with a $\mathcal{D}$ dimensional Hilbert space and operators $ \hat{\Omega}$ that
each contain $M$ elements in the configuration basis where $1\leq M \leq \mathcal{D}^{2}$. The
Behemoths form a basis in $\mathbb{O}$ but to facilitate the counting, we consider sums of Behemoths
with coefficients zero and one -- the set of operators living in $\bar{\mathbb{O}}\subset
\mathbb{O}$. We expect, however, the scalings we have found to hold for arbitrary coefficients of
order one and for any basis ``sufficiently different" from the eigenstate basis. There are then
$2^{\mathcal{D}^2}$ distinct operators in $\bar{\mathbb{O}}$. Of these, there are $\mathcal{D}^2$
Behemoth operators and $(\log \mathcal{D})^2$ physical two-point local operators.  Assuming that the
random matrix scaling is obeyed by all typical operators within each class, it follows that
super-ETH scaling is observed for
$\sum_{k=1}^{\mathcal{D}-1}\left( \begin{array}{c} \mathcal{D}^2 \\ k \end{array}\right)$
operators, ETH scaling for
$\left( \begin{array}{c} \mathcal{D}^2 \\ \mathcal{D} \end{array}\right)$
and sub-ETH scaling for the rest. For large $\mathcal{D}$ this
gives
$\exp(\mathcal{D}(\log\mathcal{D}+1))/\sqrt{\mathcal{D}}$
super-ETH operators. The sub-ETH operators appear exponentially more frequently than the rest while
physical operators are doubly exponentially suppressed again in the space of operators with
$\mathcal{D}^{-\delta}$ scaling with $\delta\geq 1/2$. From this point of view, typical operators
exhibit sub-ETH scaling while ETH scaling is exponentially rare. These scalings are compounded when
we allow for arbitrary coefficients in sums of Behemoth operators.

\begin{acknowledgments}
We thank A.~B\"{a}cker, Y.~Bar~Lev, W. Beugeling, D.~Luitz and R.~Moessner for related
collaborations and H.~Nugent and A.~Sen for helpful discussions. I.M.K. acknowledges the support of
German Research Foundation (DFG) Grant No. KH~425/1-1 and the Russian Foundation for Basic Research.
\end{acknowledgments}

\bibliography{references}

\begin{thebibliography}{52}
\bibitem [{SMS()}]{SMSubscriptsFootnote}%
  \BibitemOpen
  \href@noop {} {}\bibinfo {note} {Here and further we use for brevity
  subscripts $k=1,2,\ldots$ for $\alpha,i$ or $\beta,j$ and so on,
  respectively.}\BibitemShut {Stop}%
\end{thebibliography}

\newpage

\onecolumngrid
\newpage
\begin{center}
{\bf\Large Supplemental Materials}
\vspace{1cm}
\end{center}
\twocolumngrid

%%%%%%%%%%%%%%%%%%%%%%%%%%%%%%%%%%%%%%%%%%%%%%%%%%%%%%%
% Counter resetting for supplementaries
\setcounter{page}{1} \renewcommand{\thepage}{S\arabic{page}}

\setcounter{figure}{0}   \renewcommand{\thefigure}{S\arabic{figure}}

\setcounter{equation}{0} \renewcommand{\theequation}{S.\arabic{equation}}

\setcounter{table}{0} \renewcommand{\thetable}{S.\arabic{table}}

\setcounter{section}{0} \renewcommand{\thesection}{S.\Roman{section}}

\renewcommand{\thesubsection}{S.\Roman{section}.\Alph{subsection}}

% referring to subsections: don't prefix by section because that's already contained in
% \thesubsection.  Prefixes can be set using \p@*****.  Needs to be set within \makeatletter and
% \makeatother so that @ can be used in this way --->

\makeatletter
\renewcommand*{\p@subsection}{}
\makeatother

\renewcommand{\thesubsubsection}{S.\Roman{section}.\Alph{subsection}-\arabic{subsubsection}}

\makeatletter
\renewcommand*{\p@subsubsection}{}  % referring to subsubsections
\makeatother

%%%%%%%%%%%%%%%%%%%%%%%%%%%%%%%%%%%%%%%%%%%%%%%%%%%%%%%

%%%%%%%%%%%%%%%%%%%%%%%%%%%%%%%%%%%%%%%%%%%%%%%%%%%%%%%%%%%%%
%%%%%%%%%%%%%%%%%%%%%%%%%%%%%%%%%%%%%%%%%%%%%%%%%%%%%%%%%%%%%
%% Overview
%%%%%%%%%%%%%%%%%%%%%%%%%%%%%%%%%%%%%%%%%%%%%%%%%%%%%%%%%%%%%
%%%%%%%%%%%%%%%%%%%%%%%%%%%%%%%%%%%%%%%%%%%%%%%%%%%%%%%%%%%%%
\section{Overview}
\label{suppsec:Overview}

In the Supplemental Materials, we provide supporting information and data:

\begin{itemize}

\item We have compared our random matrix theory (RMT) predictions for distributions of matrix
  elements with numerical calculations for a number of different many-body lattice systems
  (fermionic, bosonic, magnetic).  Some of these are presented explicitly in the main text.  In
  Section \ref{suppsec:various_manybody_models} we list the different Hamiltonians which have been
  used to test the RMT predictions.  We also present numerical distributions of off-diagonal matrix
  elements for a few additional systems not shown in the main text, to further highlight the
  universal nature of our results.

\item In the main text, the focus has been on off-diagonal matrix elements, for which it is easier
  to obtain better statistics.  In Section \ref{suppsec:diagonal_data} we show examples of
  distributions of diagonal matrix elements (eigenstate expectation values), which obey the RMT
  predictions just as well.

\item In the main text we have pointed out that some Behemoth operators will show anomalous behavior
  due to energetic bias of the many-body Hamiltonian for some configurations.  We provide examples
  in Section \ref{suppsec:RMTfail}.

\item A new result reported in this work is the way correlations are manifested in the distribution
  of local operators.  By comparing random matrix eigenstates with many-body eigenstates, we further
  substantiate the finding of subtle many-body correlations present in the many-body eigenstates
  even in the middle of the many-body spectrum.  (Section \ref{suppsec:randommat}.)

\item In Section \ref{suppsec:ensembles} we provide details of derivations of the RMT predictions
  for probability distributions of Behemoths.  The main text focused on GOE systems, with one
  example for a GUE system.  Here we provide results for all three standard random matrix classes
  (GOE, GUE and GSE).

\end{itemize}

%%%%%%%%%%%%%%%%%%%%%%%%%%%%%%%%%%%%%%%%%%%%%%%%%%%%%%%%%%%%%
%%%%%%%%%%%%%%%%%%%%%%%%%%%%%%%%%%%%%%%%%%%%%%%%%%%%%%%%%%%%%
%% Atypical States
%%%%%%%%%%%%%%%%%%%%%%%%%%%%%%%%%%%%%%%%%%%%%%%%%%%%%%%%%%%%%
%%%%%%%%%%%%%%%%%%%%%%%%%%%%%%%%%%%%%%%%%%%%%%%%%%%%%%%%%%%%%

\section{Various many-body systems}
\label{suppsec:various_manybody_models}

%%%%%%%%%  FIGURE IN SUPPLEMENTARY  %%%%%%%%%%%%%%%%%%%%%%%%%%%%
\begin{figure*}[tbp]
\centering
\hspace{-.5cm}
\centering
\includegraphics[width=1.3\columnwidth]{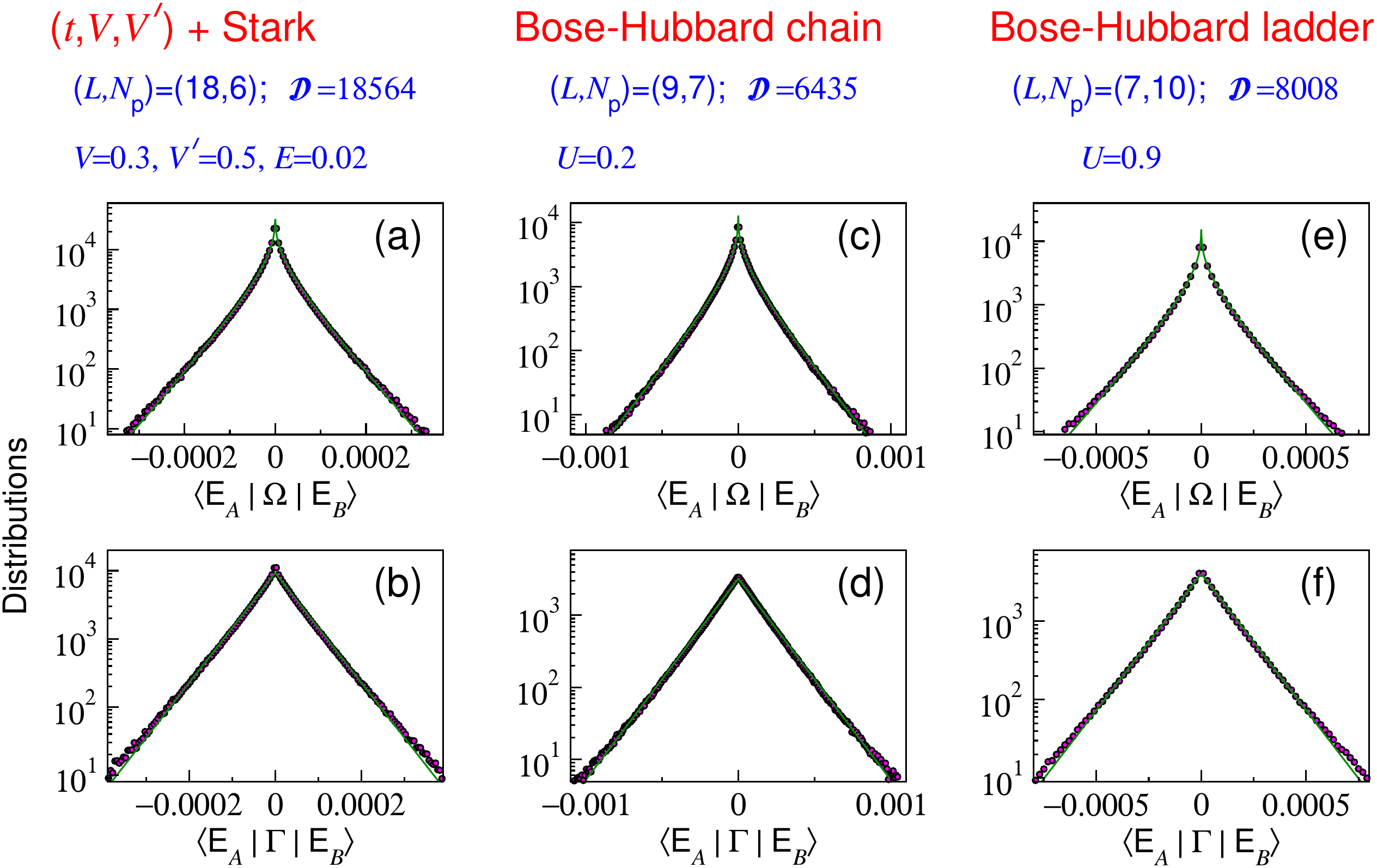}
\caption{The distributions of off-diagonal matrix elements of Behemoth operators for three different
  many-body systems, both non-hermitian (top row) and hermitian (bottom row).  In each case we take
  50-100 eigenstates in the middle of the spectrum and randomly choose 500 Behemoth operators.  For
  each Behemoth, the matrix element between each pair of distinct eigenstates is calculated.  As in
  the main text, the points are the normalized histograms of this data set, and the lines are the
  RMT predictions.
\label{fig:various_models}
}
\end{figure*}
%%%%%%%%%  FIGURE IN SUPPLEMENTARY  %%%%%%%%%%%%%%%%%%%%%%%%%%%%

The distributions for Behemoth operators that we have predicted using random matrix theory are
expected to be universal in the sense that, in any generic non-integrable many-body Hamiltonian,
they should hold for most Behemoths for eigenstates not too close to the spectral edge.

We have compared distributions of off-diagonal matrix elements of Behemoths in about half a dozen
different chaotic systems, some shown in the main text and some more shown in Figure
\ref{fig:various_models}.  In each case, we have experimented with the system sizes and fillings
($L$ and $N_{\rm p}$) as well as the coupling parameters.  We have found the conformance to the RMT
predictions to be very robust.  The obvious exceptions are when the Hamiltonian is too close to
integrability and when large interactions create non-universal (banded) structures in the spectrum.

In the main text, we have shown distributions of off-diagonal matrix elements for three different
non-integrable many-body systems. In Figure 2(a,b) of the main text, we have used the
anisotropic Heisenberg chain (XXZ chain) with both nearest-neighbor (NN) and next-nearest-neighbor
(NNN) interactions:
\begin{multline}
H ~=~ J_1 \sum_{i=1}^{L-1}  \left( S_i^+S_{i+1}^- +S_i^-S_{i+1}^+  + \Delta_1 S_i^zS_{i+1}^z \right)
\\
~+~ J_2 \sum_{i=2}^{L-2}  \left( S_i^+S_{i+2}^- +S_i^-S_{i+2}^+  + \Delta_2 S_i^zS_{i+2}^z \right)
.
\end{multline}
The summation is over the site index.  Note that the NNN coupling between sites 1 and 3 is omitted
(summation starts from $i=2$ instead of $i=1$), in order to avoid reflection symmetry.  The $J_2$
NNN coupling breaks integrability.  The integrable $XXZ$ chain, e.g., in Figure 4(c) of the main text,
is obtained for $J_2=0$.

Also in Figure 2(c) of the main text, we have shown the distribution of off-diagonal
matrix elements for a Bose-Hubbard flux ladder:
\begin{multline}
  H ~=~ -  \sum_{l} \sum_{\sigma=\mathsf{L},\mathsf{R}}
\left( b_{l;\sigma}^\dagger b_{l+1;\sigma} +  b_{l+1;\sigma}^\dagger b_{l;\sigma} \right)
\\
  -  \sum_l \left( e^{-i\phi l} b_{l;\mathsf{L}}^\dagger b_{l;\mathsf{R}}
   + e^{+i\phi l} b_{l;\mathsf{R}}^\dagger b_{l;\mathsf{L}} \right)
\\
  ~+~ \frac{U}{2} \sum_{l,\sigma} n_{l;\sigma} (n_{l;\sigma}-1) .
\label{eq:Ham_BH_fluxladder}
\end{multline}
Here $\sigma$ is the leg index taking the values $\mathsf{L}$ and $\mathsf{R}$ for the left and the
right leg.  The right leg contains one unmatched site in order to break reflection symmetries; the
geometry is shown in the same figure.  In the interaction term the site index $l$ therefore runs
from $1$ to $(L-1)/2$ for the left leg and from $1$ to $(L+1)/2$ for the right leg.  The Peierls
phases on the rungs create a flux through every plaquette; the bosons are thus subjected to a
magnetic field and the system breaks time reversal symmetry.  Accordingly, the Behemoth matrix
elements $\langle \mathsf{E}_A \left|\Omega\right|\mathsf{E}_B\rangle$ are distributed according to
our prediction for the GUE class.

In Figure 4(a,b) of the main text, we have used a fermionic tight-binding chain with both NN and NNN
interactions, and subjected this system to a Gaussian disorder:
\begin{multline}
H ~=~ - t\sum_{i=1}^{L-1} \left( c_i^{\dagger}c_{i+1} + c_{i+1}^\dagger c_{i} \right)
\\
~+~ V\sum_{i=1}^{L-1} n_{i}n_{i+1}  ~+~ V'\sum_{i=1}^{L-2} n_{i}n_{i+2}
\\ ~+~ W \sum_{i=1}^{L} \xi_i n_{i}  .
\label{eq:Ham_fermions_disorder}
\end{multline}
Here $c_i$, $c_i^{\dagger}$ are fermionic annihilation and creation operators for the $i$-th site,
respectively, $n_{i} = c_{i}^{\dagger}c_{i}$, and $\xi_{i}$ is a Gaussian random variable with mean
$0$ and variance $1$.  The hopping constant $t$ can be set to $t=1$ without loss of generality.  For
small values of the $W$, this system is chaotic and has GOE level statistics.  Accordingly, the
off-diagonal matrix elements of Behemoth operators have a distribution showing excellent agreement
with the RMT prediction, as we have shown in Figure 4(a) of the main text.

In Figure \ref{fig:various_models} we show similar results for three additional similar systems.

In panels (a,b) of Figure \ref{fig:various_models} we subject the spinless-fermion chain to a Stark
(electric or graviational) field:
\begin{multline}
H ~=~ - \sum_{i=1}^{L-1} \left( c_i^{\dagger}c_{i+1} + c_{i+1}^\dagger c_{i} \right)
\\
~+~ V\sum_{i=1}^{L-1} n_{i}n_{i+1}  ~+~ V'\sum_{i=1}^{L-2} n_{i}n_{i+2}
\\ ~+~ E \sum_{i=1}^{L} i n_{i}  .
\label{eq:Ham_fermions_Stark}
\end{multline}
The Stark field $E$ causes the sites to have uniformly increasing bare on-site energy.  A larger
value of the Stark field has a localizing effect, and very large $E$ breaks the spectrum up into
bands.  For small $E$, the system is non-integrable and has GOE level statistics.  Figure
\ref{fig:various_models}(a,b) shows that the distributions of $\langle \mathsf{E}_A
\left|\Omega_{\bf n n'}\right|\mathsf{E}_B\rangle$ and $\langle \mathsf{E}_A \left|\Gamma_{\bf n
  n'}\right|\mathsf{E}_B\rangle$ values follow the predicted distributions.

Disorder breaks reflection symmetry, as does the Stark field, so it is not necessary to modify the
Hamiltonians \eqref{eq:Ham_fermions_disorder} or \eqref{eq:Ham_fermions_Stark} in order to avoid
reflection symmetry.

In panels (c,d) of Figure \ref{fig:various_models} we consider the Bose-Hubbard Hamiltonian on a
chain:
\begin{equation}
  H ~=~ -  \sum_{i=1}^{L} t_i \left( b_{i}^\dagger b_{i+1} +  b_{i+1}^\dagger b_{i} \right)
  ~+~ \frac{U}{2} \sum_{i} n_{i} (n_{l}-1)
\label{eq:Ham_BH_chain}
\end{equation}
with $t_1=\frac{1}{2}$ and $t_{i\neq1}=1$; the hopping on the first bond is reduced to break
reflection symmetry.  Here $b_i$, $b_i^{\dagger}$ are bosonic annihilation and creation operators
for the $i$-th site, respectively.

Finally, Figure \ref{fig:various_models}(e,f) consider the GOE version of the Bose-Hubbard ladder,
i.e., interacting bosons on a ladder without flux, Eq.\ \eqref{eq:Ham_BH_fluxladder} with $\phi=0$.

\section{Diagonal Matrix Elements}
\label{suppsec:diagonal_data}

%%%%%%%%%  FIGURE IN SUPPLEMENTARY  %%%%%%%%%%%%%%%%%%%%%%%%%%%%
\begin{figure}[tbp]
\centering
\hspace{-.5cm}
\centering
\includegraphics[width=0.85\columnwidth]{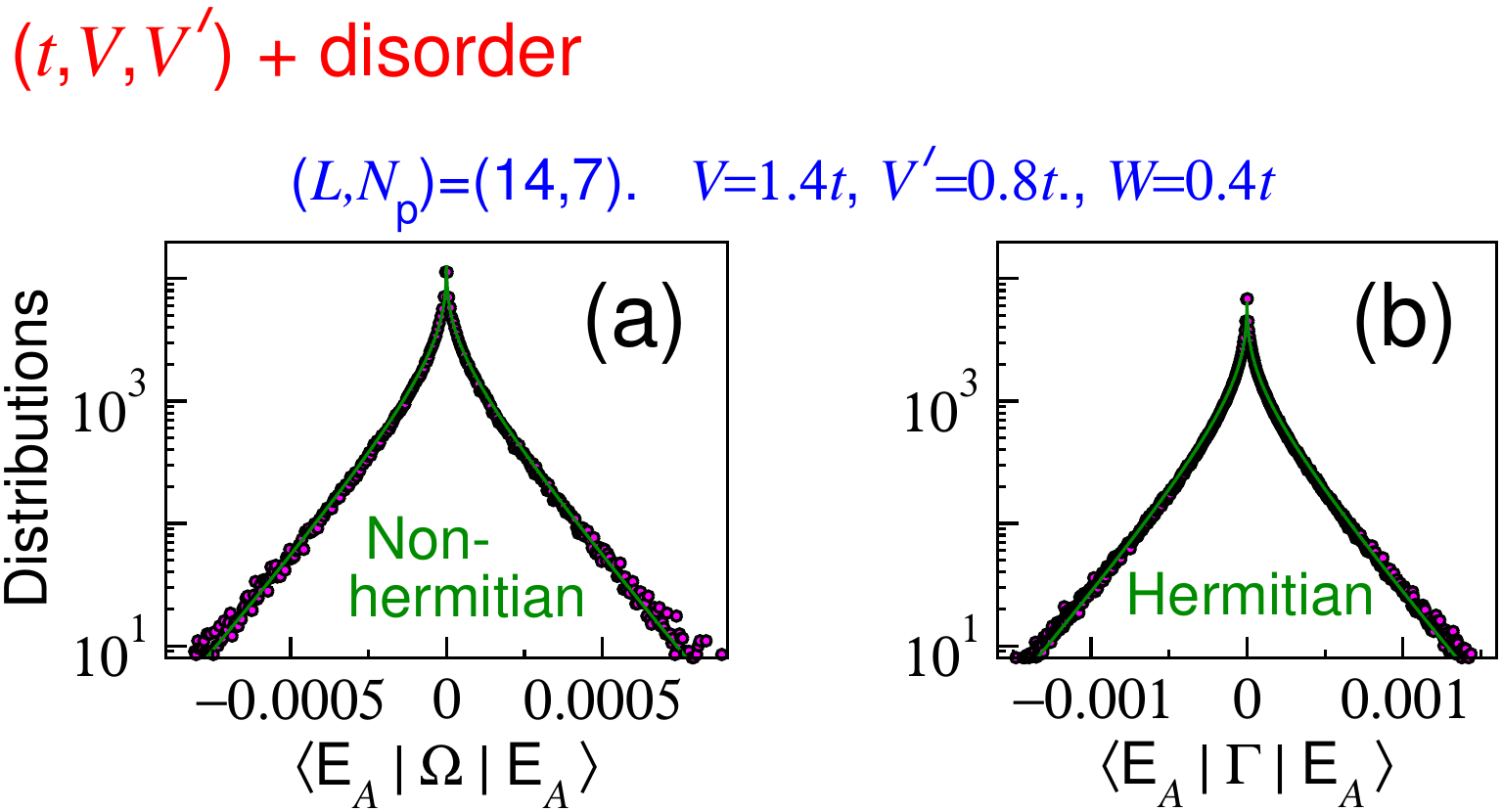}
\caption{The distributions of \emph{diagonal} matrix elements of Behemoth operators for the
  disordered spinless-fermion chain.  As elsewhere, the points are the normalized histograms of
  many-body data, and the lines are RMT predictions.
\label{fig:diagonal}
}
\end{figure}
%%%%%%%%%  FIGURE IN SUPPLEMENTARY  %%%%%%%%%%%%%%%%%%%%%%%%%%%%

In the main text, we presented only distributions of off-diagonal matrix elements.  This is
convenient for gathering statistically significant datasets because there are more off-diagonal
matrix elements than diagonal matrix elements.

In Figure \ref{fig:diagonal}, we show comparisons for the distributions of diagonal matrix elements.
In this case we have used the disordered chain, Eq.\ \eqref{eq:Ham_fermions_disorder}, and collected
data for several disorder realizations to obtain better statistics.  In contrast to the off-diagonal
case, the hermitian and non-hermitian operators now have the same distribution shape (both $K_0$) but different
widths (Section \ref{suppsec:ensembles}; Eqs.\ \eqref{eq:P_diag_herm(y)_GOE} and
\eqref{eq:P_x_GOE}).

%%%%%%%%%%%%%%%%%%%%%%%%%%%%%%%%%%%%%%%%%%%%%%%%%%%%%%%%%%%%%
%%%%%%%%%%%%%%%%%%%%%%%%%%%%%%%%%%%%%%%%%%%%%%%%%%%%%%%%%%%%%
%% Atypical States
%%%%%%%%%%%%%%%%%%%%%%%%%%%%%%%%%%%%%%%%%%%%%%%%%%%%%%%%%%%%%
%%%%%%%%%%%%%%%%%%%%%%%%%%%%%%%%%%%%%%%%%%%%%%%%%%%%%%%%%%%%%
\section{Atypical Operators}
\label{suppsec:RMTfail}

In the main text, we have reported that the random matrix theory predictions for distributions can
be violated for Behemoth operators $\Omega_{\boldsymbol{n}\boldsymbol{n'}} =
\ket{\boldsymbol{n}}\bra{\boldsymbol{n'}}$ if one or both of configurations $\boldsymbol{n}$ and
$\boldsymbol{n'}$ are energetically penalized (or favored) by the Hamiltonian.  The same is true for
the hermitian Behemoth $\Gamma_{\boldsymbol{n}\boldsymbol{n'}} =
\ket{\boldsymbol{n}}\bra{\boldsymbol{n'}} + \ket{\boldsymbol{n'}}\bra{\boldsymbol{n}}$.  We present
an explicit example in Figure \ref{fig:energy}.

In this case, the special configuration is $\ket{\boldsymbol{n}} = \ket{000000001111111}$, a
 ferromagnetic configuration for this filling containing one domain wall.  The
Hamiltonian couplings used are antiferromagnetic (both NN and NNN).  Due to this physics, the
configuration $\ket{\boldsymbol{n}}$ is energetically penalized, i.e., the amplitudes of eigenstates
in this configuration are high for the highest-energy eigenstates and therefore small for other
eigenstates by normalization.  This is shown in panel (b) of Figure \ref{fig:energy}.

The distribution of matrix elements of $\Gamma_{\boldsymbol{n}\boldsymbol{n'}}$ between pairs of
eigenstates in the middle of the spectrum is shown in Figure \ref{fig:energy}(a).  The distribution
is much narrower than that predicted by random matrix theory, i.e., the values of $\langle
\mathsf{E}_A \left|\Omega \right| \mathsf{E}_B \rangle$ are much smaller on average than the RMT
prediction.  This follows directly from the fact that the weights of $\boldsymbol{n}$ are
anomalously small in these eigenstates.  If one chooses both $\boldsymbol{n}$ and $\boldsymbol{n'}$
to be atypical in this way, the distribution turns out to be even narrower.

Interestingly, the form of the distribution --- exponential for the hermitian Behemoths and Bessel
($K_0$) for the non-hermitian Behemoths --- is still obeyed, but with a modified width.  This
implies that the coefficients of  $\ket{\boldsymbol{n}}$ in eigenstates in the middle of the
spectrum are, while anomalously small, still approximately Gaussian-distributed.

%%%%%%%%%  FIGURE IN SUPPLEMENTARY %%%%%%%%%%%%%%%%%%%%%%%%%%%%
\begin{figure}[tbp]
\centering
\hspace{-.5cm}
\centering
\includegraphics[width=0.95\columnwidth]{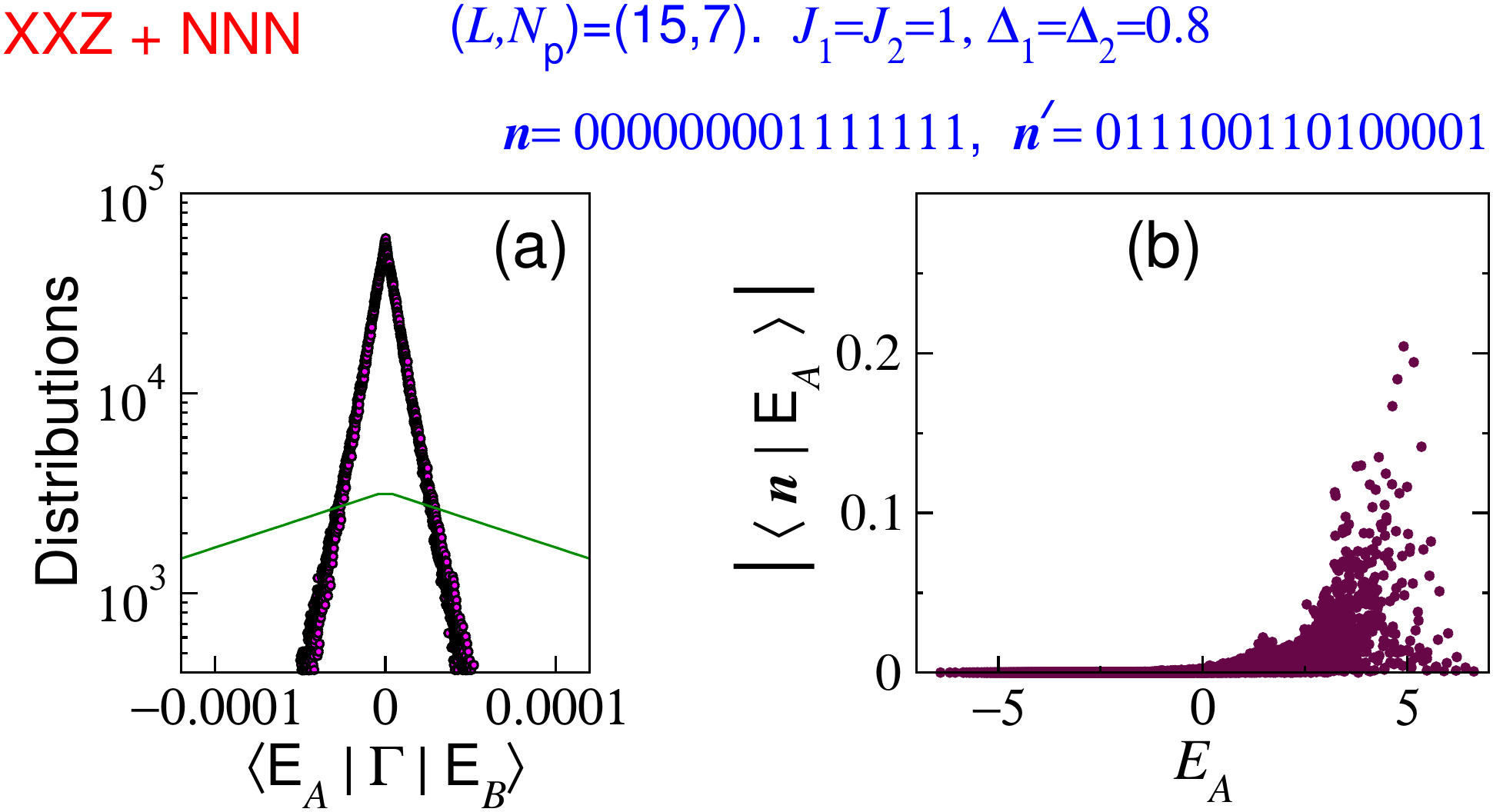}
\caption{An example of an atypical Behemoth operator, whose matrix elements do not follow the
  distribution predicted using random matrix theory.
The configurations $\boldsymbol{n}$ and $\boldsymbol{n}'$ used to build the operator
$\Gamma_{\boldsymbol{n}\boldsymbol{n'}} = \ket{\boldsymbol{n}}\bra{\boldsymbol{n'}} +
\ket{\boldsymbol{n'}}\bra{\boldsymbol{n}}$ are indicated near the top.
(a) Distributions.  As in other figures, the green line is the RMT prediction and the dots are
appropriately normalized histograms drawn from many-body data.  The central 20\% of the eigenstates
are used.
(b) Weight of the configuration $\ket{\boldsymbol{n}} = \ket{000000001111111}$ in all eigenstates,
shown as a function of eigenenergy.  The configuration is energetically penalized: only
very-high-energy eigenstates have significant weight.  This is responsible for the anomalous
behavior of the Behemoth.
\label{fig:energy}
}
\end{figure}
%%%%%%%%%  FIGURE IN SUPPLEMENTARY %%%%%%%%%%%%%%%%%%%%%%%%%%%%

%%%%%%%%%%%%%%%%%%%%%%%%%%%%%%%%%%%%%%%%%%%%%%%%%%%%%%%%%%%%%
%%%%%%%%%%%%%%%%%%%%%%%%%%%%%%%%%%%%%%%%%%%%%%%%%%%%%%%%%%%%%
%% Operator Randomness
%%%%%%%%%%%%%%%%%%%%%%%%%%%%%%%%%%%%%%%%%%%%%%%%%%%%%%%%%%%%%
%%%%%%%%%%%%%%%%%%%%%%%%%%%%%%%%%%%%%%%%%%%%%%%%%%%%%%%%%%%%%
\section{Eigenstate Correlations and Local Operators}
\label{suppsec:randommat}

%%%%%%%%%  FIGURE IN SUPPLEMENTARY %%%%%%%%%%%%%%%%%%%%%%%%%%%%
\begin{figure*}[tbp]
\begin{center}
\includegraphics[width=1.3\columnwidth]{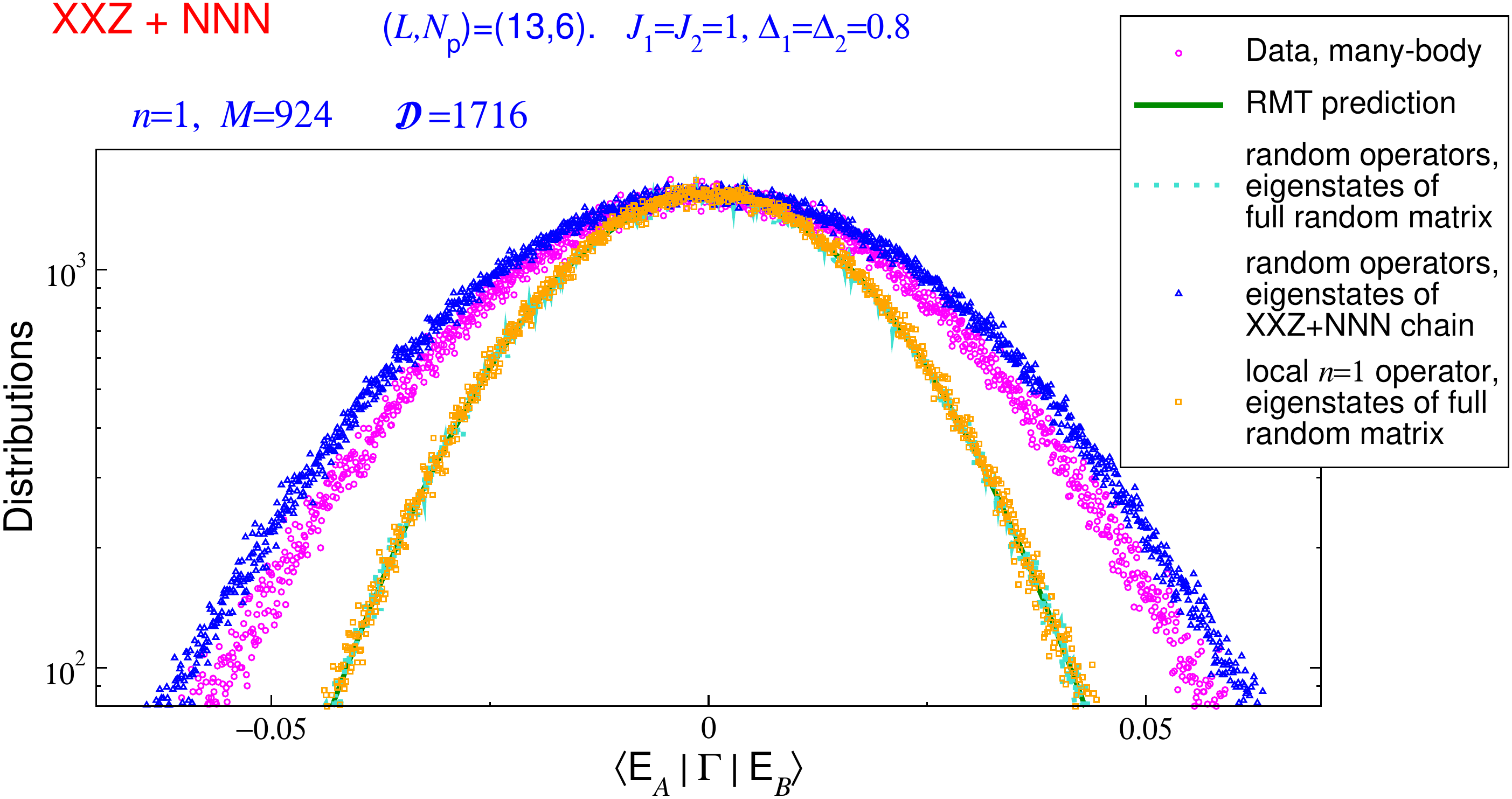}
\caption{Comparison of distributions of off-diagonal matrix elements, for local operators and random
  operators with the same $M$, in eigenstates of the non-integrable spin chain and in eigenstates of
  a full random matrix of the same size.
As discussed in the main text, the distribution for the local operator in the spin chain eigenstates
has a larger width than the RMT prediction.  The random operator distribution is found to also have
a larger width in the spin chain eigenstates.  In the eigenstates of a full random matrix (GOE
matrix), both operators have distributions following the RMT prediction.
This comparison shows that the departure from random matrix predictions is a property of the
many-body eigenstates that is not tied to the spatial locality of the
operators.} \label{fig:randommat}
\end{center}
%\vspace{-25pt}
\end{figure*}
%%%%%%%%%  FIGURE IN SUPPLEMENTARY %%%%%%%%%%%%%%%%%%%%%%%%%%%%

In the main text, we showed that correlation effects in many-body systems are manifested in the
distributions of local operators through deviations from the RMT prediction.  The distribution of local
operators in many-body systems matches the functional form and $\mathcal{D}^{-1/2}$ scaling predicted
by RMT using the central limit theorem; however the width is somewhat larger than the RMT prediction.
In other words, correlation effects result in a partial violation of the central limit theorem.

The origin of the discrepancy lies in the eigenstates of many-body systems not being totally random.
This can be seen through the comparison in Figure \ref{fig:randommat}.  When we use an operator that
has as many nonzero entries as the nonlocal operator but whose entries are randomly chosen among the
matrix elements, then the same type of discrepancy is observed.  On the other hand, if we replace
the many-body eigenstates by the eigenstates of a full random matrix (a GOE matrix), this results in
the off-diagonal matrix element distribution following the RMT prediction, no matter which
$M=O(\mathcal{D})$ operator is used.

This shows that the correlations (non-randomness) reside in the eigenstate structure and do not
depend much on exactly which $M=O(\mathcal{D})$ operator is used.

%%%%%%%%%%%%%%%%%%%%%%%%%%%%%%%%%%%%%%%%%%%%%%%%%%%%%%%%%%%%%
%%%%%%%%%%%%%%%%%%%%%%%%%%%%%%%%%%%%%%%%%%%%%%%%%%%%%%%%%%%%%
%% GUE and GSE
%%%%%%%%%%%%%%%%%%%%%%%%%%%%%%%%%%%%%%%%%%%%%%%%%%%%%%%%%%%%%
%%%%%%%%%%%%%%%%%%%%%%%%%%%%%%%%%%%%%%%%%%%%%%%%%%%%%%%%%%%%%
\section{Results for Gaussian Ensembles}
\label{suppsec:ensembles}

In this section we present the random matrix theory derivations for the distributions of matrix
elements of Behemoths.  We consider the three common ensembles of random matrices: namely the
Gaussian Orthogonal, Gaussian Unitary, and Gaussian Symplectic Ensembles (GOE, GUE and GSE
respectively).

We consider eigenvectors $\ket{E_{\alpha}}$ of a random matrix with coefficients
\begin{equation}\label{eq:RMT_En_expansion}
\la i| E_\alpha\ra = u_{\alpha,i} \ ,
\end{equation}
where $i$ is the basis index.  We use the interpretation that the random matrix represents a
single-particle hopping Hamiltonian on a fully connected graph.  The basis indices can therefore be
referred to as site indices.  The objects of interest in this work are the inter-site hopping
operators
\begin{equation}\label{eq:RMT_A_operator}
\hat \omega_{ij} = \hat d_i^\dagger \hat d_j \equiv \left|i\ra\la j \right|  \ .
\end{equation}
Here $\hat d_i^\dagger$ ($\hat d_i$) is the single-particle creation (annihilation) operator at site
$i$.  In the many-body interpretation, these operators correspond to the Behemoths, whose matrix
elements are the subject of this work.  Although the name Behemoth arises in the many-body
interpretation, below we will refer to the random matrix operators $\hat \omega_{ij}$ as Behemoth
operators.

We are interested in the distributions of matrix elements of $\hat \omega_{ij}$ in the eigenstates
$\ket{E_{\alpha}}$, i.e., in the distributions $P_\omega(\omega_{nm}^{ij})$ of
\begin{equation}\label{eq:RMT_Xmn_ij}
\omega_{ij}^{\alpha\beta}=\la E_\alpha | \hat d_i^\dagger \hat d_j | E_\beta\ra = u_{\alpha,i}^\star
u_{\beta,j} \ .
\end{equation}
We also consider  the distributions $P_\gamma(\gamma_{nm}^{ij})$ of matrix elements of the
Hermitian version, i.e., of
\begin{equation}\label{eq:RMT_Ymn_ij_herm}
\gamma_{ij}^{\alpha\beta} = \la E_\alpha| d_i^\dagger d_j + d_j^\dagger d_i |E_\beta\ra =
u_{\alpha,i}^\star u_{\beta,j}+u_{\alpha,j}^\star u_{\beta,i}  \ .
\end{equation}
The many-body interpretation of these objects depend significantly on site ($i,j$) and eigenstate
($\alpha,\beta$) indices.  There are a few cases:
\begin{itemize}%[leftmargin=0mm]
\item $i\neq j$, $\alpha\neq\beta$.  This gives off-diagonal matrix elements of the Behemoth
  operators.  In this case $\gamma_{ij}^{\alpha\beta} =
  \omega_{ij}^{\alpha\beta}+\omega_{ji}^{\alpha\beta}$.  This is the case mainly focused on in this
  work for many-body systems.
\item $i\neq j$, $\alpha=\beta$.   This gives diagonal matrix elements (eigenstate expectation
  values) of Behemoths.  In this case $\gamma_{ij}^{\alpha\alpha} = 2\Re
  \omega_{ij}^{\alpha\alpha}$.
\item We can also consider the case $i=j$, i.e., operators $\omega_{ii} = \hat d_i^\dagger \hat d_i$
  in the many-body language.  In the many-body analogy, these represent
  $\Omega_{\boldsymbol{n}\boldsymbol{n}} = \vert\boldsymbol{n}\rangle\langle \boldsymbol{n}\vert$,
  projectors onto many-body configurations.  For completeness we provide some results on the
  distributions of these operators.  Since  $\omega_{ii}$ is hermitian by construction,
  $\gamma_{ii}^{\alpha\beta} = 2\omega_{ii}^{\alpha\beta}$.
\end{itemize}

In the limit $N\to \infty$ the real-valued components of each $u_{\alpha,i}$ can be approximated by
independent identically distributed gaussian random variables with zero mean (see, e.g.,
\cite{beenakker1997random,mehta2004random} and references therein).  The variance of these real-valued components $\sigma_\beta^2$ is governed by the ``unitarity'' condition
\begin{equation}\label{eq:u_unitarity}
\sum_i u_{\alpha,i}^\star u_{\beta,i} = \delta_{\alpha,\beta}, \quad
\sum_\alpha u_{\alpha,i}^\star u_{\alpha,j} = \delta_{i,j},
\end{equation}
as
\begin{equation}\label{eq:sigma_beta}
1 = \sum_i |u_{\alpha,i}|^2 \approx N\la |u_{\alpha,i}|^2\ra = N\beta\sigma_\beta^2 \ , \quad \sigma_\beta^2 = 1/(\beta N)
\end{equation}
with $\beta = 1$, $2$, $4$ being the number of real-valued components of each matrix element $u_{\alpha,i}$ for GOE, GUE, and GSE, respectively.

%%%%%%%%%%%%%%%%%%%%%%%%%%%%%%%%%%%%%%%%%%%%%%%%%%%%
%                   GOE
%%%%%%%%%%%%%%%%%%%%%%%%%%%%%%%%%%%%%%%%%%%%%%%%%%%%

\subsection{GOE}

Eigenvectors of GOE matrices have real coefficients.  According to the above-mentioned approximation
the real coefficients $u_{\alpha,i}$ are independent gaussian variables with zero mean and variance
$\sigma_1^2 = 1/N$:
\begin{equation}
P_u(u) = \frac{e^{-u^2/2\sigma_1^2}}{\sqrt{2\pi \sigma_1^2}} \ .
\end{equation}
The distribution of $\omega_{ij}^{\alpha\beta}$ is then
\begin{multline}\label{eq:P_x_GOE}
P_\omega(x) = \int_{-\infty}^\infty du_1 du_2 P_u(u_1) P_u(u_2) \delta(x - u_1 u_2) \\ =
2\int_{0}^\infty du_1 P_u(u_1) P_u(x/u_1) \int_{-\infty}^\infty \delta(x - u_1 u_2) du_2
\\ =
2\int_{0}^\infty \frac{du_1}{2\pi \sigma_1^2 u_1} \exp\lb-\frac{u_1^2+(x/u_1)^2}{2\sigma_1^2}\rb
\\ =
\frac{1}{\pi \sigma_1^2} K_0\lp \frac{|x|}{\sigma_1^2}\rp \ ,
\end{multline}
where $K_0(z)$ is the modified Bessel function of the second kind.
This result is quoted in the main text for the off-diagonal matrix elements of non-hermitian
Behemoths.

The above calculation is unchanged for the case of \emph{diagonal} matrix elements of non-hermitian
Behemoths ($\omega_{ij}^{\alpha\alpha} = u_{\alpha,i}u_{\alpha,j}$), which thus have the same
distribution $P_\omega(x)$.

We now turn to the hermitian operators $\hat\gamma_{ij}$.
To calculate the off-diagonal matrix elements \eqref{eq:RMT_Ymn_ij_herm} of $\hat\gamma_{ij}$ one
should consider the convolution of two distributions of the form of \eqref{eq:P_x_GOE}.  Using the
Fourier transform of \eqref{eq:P_x_GOE}
\begin{equation}\label{eq:P(q)_FT_GOE}
\tilde P_\omega(q) = \int_{-\infty}^\infty e^{i\omega q} P_\omega(\omega) d\omega = \frac{1}{\lb
  1+\sigma_1^4 q^2 \rb^{1/2}} \ ,
\end{equation}
one can calculate the Fourier transform of the $\gamma$-distribution
\begin{equation}
\tilde P_{\gamma}(q) = \lb\tilde P_\omega(q)\rb^2 = \frac{1}{1+\sigma_1^4q^2} \ ,
\end{equation}
which leads to
\begin{equation}\label{eq:P(y)_GOE}
P_\gamma(y) = \frac{1}{2\pi}\int_{-\infty}^\infty e^{-i q y}  \frac{dq}{1+\sigma_1^4q^2} = \frac{e^{-|y|/\sigma_1^2}}{2\sigma_1^2} \ .
\end{equation}

As all entries $u_{\alpha,i}$ are real, the \emph{diagonal} matrix elements of the hermitian
operator $\hat\gamma_{ij}$ are given by $\gamma_{ij}^{\alpha\alpha}=2\omega_{ij}^{\alpha\alpha}$.
Their distribution is thus
\begin{equation}\label{eq:P_diag_herm(y)_GOE}
P_{\gamma,\rm diag}(y) = P_\omega(y/2)
\end{equation}
where $P_\omega$ is defined in \eqref{eq:P_x_GOE}.

\subsubsection{Sums of matrix elements}

We now consider the distribution of the sum ${\omega_M} = \sum_k \omega_k$ of $M$ independent
non-hermitian Behemoths $\omega_k$
\cite{SMSubscriptsFootnote}.
One can calculate this analogously to Eq.~\eqref{eq:P(y)_GOE} using the Fourier transform
\eqref{eq:P(q)_FT_GOE}:
\begin{multline}\label{eq:P(x_M)}
P_{{\omega_M}}(X) = \frac{1}{2\pi}\int_{-\infty}^\infty e^{-i q X}\lb\tilde P_\omega(q)\rb^{M}{d q}
\\
= \frac{1}{2\pi}\int_{-\infty}^\infty e^{-i q X}  \frac{d q}{\lb 1+\sigma_1^4 q^2\rb^{M/2}}
\\
= \frac{\lp{|X|}/{2\sigma_1^2}\rp^{\frac{M-1}2}}{\sqrt{\pi}\Gamma[M/2] \sigma_1^{2}} K_{\frac{1-M}2}\lp \frac{|X|}{\sigma_1^2}\rp
\\
\equiv \bar P_{M,\sigma_1} (X)
 \ .
\end{multline}
From now on, we use the notation $\bar P_{M,\sigma_1} (X)$ for this $M$-distribution.

The result \eqref{eq:P(x_M)} has been quoted in the main text in the context of building local
operators out of Behemoths.
For large $M\gg 1$ this distribution can be approximated by the gaussian distribution
\begin{equation}\label{eq:P_sumX_GOE}
P_{\omega_M}(X)\approx \frac{e^{-X^2/(2 M \sigma_1^4)}}{\sqrt{2\pi M}\sigma_1^2} \ ,
\end{equation}
with the variance scaling as $\sigma^2 =M \sigma_1^4 \sim M N^{-2}$.

For sums of non-hermitian operators, the distribution is the same for both off-diagonal and for
diagonal matrix elements, given by Eq.\ \eqref{eq:P(x_M)}.

Analogously, for the sum $\gamma_M = \sum_k \gamma_k$ of hermitian operators
\eqref{eq:RMT_Ymn_ij_herm} one obtains
\begin{equation}\label{eq:P(y_M)_offdiag}
P_{{\gamma_M}}(Y) = \bar P_{2M,\sigma_1} (Y)  \
\end{equation}
for off-diagonal matrix elements, $\alpha\ne\beta$, and
\begin{equation}\label{eq:P(y_M)_diag}
P_{{\gamma_M,\rm diag}}(Y) = \bar P_{M,\sigma_1} (Y/2)  \
\end{equation}
for diagonal ones $\alpha=\beta$.
%The variance of this distribution scales as $\sigma^2 =(2M) \sigma_1^4 \sim 2 M / N^{2}$.

Note that Eq.~\eqref{eq:P(x_M)} also applies to the off-diagonal elements $\alpha\ne\beta$ of the
projection operators, $i=j$, for all $M$ and for diagonal elements $\alpha=\beta$ of these operators
for $M\ll N$.  In the case of $\alpha=\beta$ and $M\simeq N$ one needs to take into account the
finite mean value of the operator due to Eq.\ \eqref{eq:u_unitarity}.

%%%%%%%%%%%%%%%%%%%%%%%%%%%%%%%%%%%%%%%%%%%%%%%%%%%%
%                   GUE
%%%%%%%%%%%%%%%%%%%%%%%%%%%%%%%%%%%%%%%%%%%%%%%%%%%%

\subsection{GUE}

Eigenstates of GUE matrices have complex coefficients in general.  Each coefficient takes the form
$u_{\alpha,i}=u'_{\alpha,i}+iu''_{\alpha,i}$ with components $u'_{\alpha,i}$ and $u''_{\alpha,i}$
that are real-valued and (in the limit $N\to \infty$) independent gaussian variables with zero mean
and variance $\sigma_{2}^{2}=1/(2N)$ \cite{beenakker1997random,mehta2004random}. The distribution
$P_u(u'_{\alpha,i},u''_{\alpha,i})$ is factorized in terms of real and imaginary parts
\begin{gather}
P_u(u',u'') = \frac{1}{2\pi \sigma_2^2} \exp\left( - \frac{u'^2 + u''^2}{2\sigma_2^2}  \right)
\end{gather}
as well as in polar coordinates $u_{\alpha,i} \equiv \rho_{\alpha,i} e^{i\theta_{\alpha,i}}$,
$P_u(\rho,\theta)d\rho d\theta \equiv P_u(u',u'')du' du''$
\begin{gather}
P_u(\rho,\theta) = \frac{\rho}{2\pi \sigma_2^2} \exp\left( - \frac{\rho^2}{2\sigma_2^2}  \right)
\end{gather}
As a result the distribution of the Behemoth operator $\omega_{ij}^{\alpha\beta}\equiv \omega_{12}=u_1^\star u_2$
\cite{SMSubscriptsFootnote}.
is also factorized $P_\omega (\rho_\omega , \theta_\omega)= P_\rho (\rho_\omega)
P_{\theta}(\theta_\omega)$ in polar coordinates, $\omega_{12}\equiv \rho_\omega e^{i\theta_\omega}$,
and can be calculated using $\rho_\omega = \rho_1\rho_2$ and $\theta_\omega = \theta_2 -
\theta_1$. The distribution of the phase is uniform
\begin{equation}
P_{\theta}(\theta_\omega) = \int \frac{d\theta_1 d\theta_2}{(2\pi)^2} \delta(\theta_\omega -
\theta_2 + \theta_1) = \frac{1}{2\pi} \ .
\end{equation}
The distribution of the amplitude $\rho_\omega$ can be calculated similarly as \eqref{eq:P_x_GOE};
the differences are in the Jacobian $\rho_\omega$, normalization coefficient, and the definition of
the width $\sigma_\beta$ \eqref{eq:sigma_beta}:
\begin{multline}
P_\rho (\rho_\omega) =  \int_{0}^\infty  \frac{d\rho_1 d\rho_2}{\sigma_2^4} e^{-(\rho_1^2+\rho_2^2)/2\sigma_2^2}\rho_1\rho_2\delta(\rho_\omega - \rho_1 \rho_2) \\
= \frac{\rho_\omega}{\sigma_2^4 }\int_{0}^\infty \frac{d\rho_1}{\rho_1} \exp\left(-\frac{\rho_1^2+(\rho_\omega/\rho_1)^2}{2\sigma_2^2}\right) \\
= \frac{\rho_\omega}{\sigma_2^4} K_0\left( \frac{\rho_\omega}{\sigma_2^2}\right)=\frac{\pi\rho_\omega}{\sigma_2^2}\bar P_{1,\sigma_2}(\rho_\omega).
\end{multline}
This distribution has been quoted in the main text and compared with data from a many-body system
with broken time reversal symmetry.

Note that the distribution $P_\omega (\rho_\omega , \theta_\omega)$ is not factorized in terms of
real and imaginary parts of $\omega_{12} = \omega'+i\omega''$.  However the marginal distributions
of $\omega'$ and $\omega''$ are identical and coincide with the distribution \eqref{eq:P(y)_GOE},
of the hermitian operator $\gamma_{ij}^{\alpha\beta}$ of the GOE case with $\sigma_1$ substituted for
$\sigma_2$
\begin{equation}
P_{\omega'}(x) = P_{\omega''}(x) = \frac{e^{-\vert x\vert/\sigma_2^2}}{2\sigma_2^2}=\bar
P_{2,\sigma_2}(x) \ .
\label{eq:P(x)prime_GUE}
\end{equation}

The diagonal matrix elements of the non-hermitian Behemoths have the same distributions as the
off-diagonal matrix elements outlined above.

We now consider the hermitian operators.

For the off-diagonal elements $i\ne j$, $\alpha\ne\beta$ we have
\begin{align*}
\gamma_{ij}^{\alpha\beta} & = \gamma'+i \gamma'' \\ & =
u_{\alpha,i}'u_{\beta,j}'+u_{\alpha,i}''u_{\beta,j}''+u_{\alpha,j}'u_{\beta,i}'+u_{\alpha,j}''u_{\beta,i}'' \\ & +
i(u_{\alpha,i}''u_{\beta,j}'-u_{\alpha,i}'u_{\beta,j}''+u_{\alpha,j}''u_{\beta,i}'-u_{\alpha,j}'u_{\beta,i}'')
\ .
\end{align*}
So the marginal distributions of $\gamma'$ and $\gamma''$ are given by \eqref{eq:P(x_M)} with $M=4$
and $\sigma_1$ replaced by $\sigma_2$,
\begin{equation}\label{eq:P(y)off-diag_GUE}
P_{\gamma'}(y) = P_{\gamma''}(y) = \bar P_{4,\sigma_2}(y) \ .
\end{equation}

The diagonal matrix elements $\alpha=\beta$ for the hermitian operator are real:
$\gamma_{ij}^{\alpha\alpha} = 2{\Re} \omega_{ij}^{\alpha\alpha} = 2{\omega_{ij}^{\alpha\alpha}}'$.
The distribution is given by Eq. \eqref{eq:P(x)prime_GUE}:
\begin{equation}
P_{\gamma,\rm diag}(y) = \frac{e^{-\vert y\vert/2\sigma_2^2}}{4\sigma_2^2}=\bar P_{2,\sigma_2}(y/2) \ ,
\end{equation}

\subsubsection{Sums of matrix elements}

We now consider the sum $\omega_M = \sum_k \omega_k$ of $M$ independent Behemoths in the GUE case.
Analogously to the single Behemoth, the marginal distributions of the real $\omega_M'$ and imaginary
$\omega_M''$ parts of $\omega_M =\omega_M'+i\omega_M''$ are identical, although the joint
distribution $P_{\omega_M}(\omega_M)$ is not factorized as
$P_{\omega_M'}(\omega_M')P_{\omega_M''}(\omega_M'')$.  The marginal distributions are given by
\eqref{eq:P(x_M)} with $M$ substituted for $2M$ and  $\sigma_1$ replaced with $\sigma_2$:
\begin{multline}
P_{\omega_M'}(X) = P_{\omega_M''}(X)
\\ =
\prod_{k=1}^M \int_{-\infty}^\infty P_\omega(\omega_k)d\omega_k'd\omega_k'' \delta\left(X-\sum_k \omega_k'\right)
\\ =
\bar P_{2M,\sigma_2}(X) \ ,
\end{multline}
These distributions approach the gaussian distribution with the variance scaling as $\sigma^2 = 2 M
\sigma_2^4 \sim M /2 N^{2}$.

%%%%%%%%%%%%%%%%%%%%%%%%%%%%%%%%%%%%%%%%%%%%%%%%%%%%
%                   GSE
%%%%%%%%%%%%%%%%%%%%%%%%%%%%%%%%%%%%%%%%%%%%%%%%%%%%

\subsection{GSE}

A matrix element of an eigenstate $\vert E_\alpha \rangle$ within the GSE ensemble takes the form
$u_{\alpha,i}\equiv u_1 = u^{(0)}_1 + \mathfrak{i} u^{(1)}_1+\mathfrak{j} u^{(2)}_1+\mathfrak{k} u^{(3)}_1$ \cite{SMSubscriptsFootnote} with imaginary units $\mathfrak{i}^2 = \mathfrak{j}^2 = \mathfrak{k}^2 = -1$, and within the same gaussian approximation
\cite{beenakker1997random,mehta2004random}
real-valued parameters $u^{(m)}_k$ are independent gaussian variables with the zero mean and
the variance $\sigma_4^2 = 1/(4N)$
\begin{equation}\label{eq:P_u_GSE}
P_u(u^{(0)}_k,u^{(1)}_k,u^{(2)}_k,u^{(3)}_k) = \frac{e^{-\sum_{m=0}^3 [u^{(m)}_k]^2/2\sigma_4^2}}{(2\pi \sigma_4^2)^2} \ .
\end{equation}

Analogously to the previous section the distribution \eqref{eq:P_u_GSE} factorizes in spherical coordinates
$u_{k} = \rho_k e^{\theta_{k}^{(0)} ({\bf I \cdot v_k})}$, 
${\bf I} = (\mathfrak{i}, \mathfrak{j}, \mathfrak{k})$, 
${\bf v}_k = \left(\cos\theta_{k}^{(1)}, \sin\theta_{k}^{(1)}\cos\theta_{k}^{(2)}, \sin\theta_{k}^{(1)}\sin\theta_{k}^{(2)})\right)$,
$0\leq\theta_{k}^{(0)},\theta_{k}^{(1)}<\pi$, $0\leq\theta_{k}^{(2)}<2\pi$,
\begin{equation}
P_u(\rho_k,\theta_k^{(0)},\theta_k^{(1)},\theta_k^{(2)})%d\rho d\theta_1d\theta_2d\theta_3
= \frac{e^{-\rho_k^2/2\sigma_4^2}\rho_k^3}{2\sigma_4^4}\frac{\sin^2\theta_k^{(0)}}{\pi/2}\frac{\sin\theta_k^{(1)} }{2}\frac{1}{2\pi}  \ .
\end{equation}

Then the distribution of $\omega = \rho_\omega e^{\theta_{\omega}^{(0)} ({\bf I \cdot v_\omega})}$ is also factorized in the corresponding
spherical coordinates $\rho_\omega = \rho_1 \rho_2$, $\theta_{\omega}^{(0)} = \theta_{1}^{(0)} = \theta_{2}^{(0)}$, ${\bf v}_\omega = {\bf v}_1-{\bf v}_2$,
\begin{equation}\label{eq:P(omega)_GSE}
P_\omega(\rho_\omega,\theta_{\omega}^{(0)},\theta_{\omega}^{(1)},\theta_{\omega}^{(2)}) =  P_{\rho}(\rho_\omega)P_\theta(\theta_{\omega}^{(0)},\theta_{\omega}^{(1)},\theta_{\omega}^{(0)}) \ ,
\end{equation}
with the homogeneous distribution of the unit vector $\omega/\rho_\omega$ over the $3$-sphere
\begin{equation}
P_\theta(\theta_{\omega}^{(0)},\theta_{\omega}^{(1)},\theta_{\omega}^{(2)}) =  \frac{\sin^2\theta_{\omega}^{(0)}}{\pi/2}\frac{\sin\theta_{\omega}^{(1)} }{2}\frac{1}{2\pi}\ ,
\end{equation}
and the amplitude distribution different from \eqref{eq:P_x_GOE} only by the Jacobian $\rho_\omega^3$, the normalization coefficient, and the definition of the width $\sigma_\beta$ \eqref{eq:sigma_beta}
\begin{multline}\label{eq:P(rho)_GSE}
P_{\rho}(\rho_\omega) = \int_{0}^\infty  \frac{d\rho_1 d\rho_2}{(2\sigma_4^4)^2} e^{-(\rho_1^2+\rho_2^2)/2\sigma_4^2}\rho_1^3\rho_2^3\delta(\rho_\omega - \rho_1 \rho_2) \\
= \frac{\rho_\omega^3}{4\sigma_4^8}\int_{0}^\infty \frac{d\rho_1}{\rho_1} \exp\left(-\frac{\rho_1^2+(\rho_\omega/\rho_1)^2}{2\sigma_4^2}\right) \\
= \frac{\rho_\omega^3}{4\sigma_4^8} K_0\left( \frac{\rho_\omega}{\sigma_4^2}\right)\equiv
\frac{\pi \rho_\omega^3}{4\sigma_4^6} \bar P_{1,\sigma_4}\left( \rho_\omega\right) \ .
\end{multline}
To calculate $P_\theta(\theta_\omega)$ we used the fact that the distribution of a vector $u_2/\rho_2$ on a unit $3$-sphere is invariant under rotation $u_2/\rho_2 \to \omega/\rho_\omega = (u_2/\rho_2)(u_1^\star/\rho_1)$.
Here we also used the cartesian to spherical coordinate transformation
\begin{subequations}
\begin{align}
u^{(0)}_k &= \rho_k \cos\theta_{k}^{(0)} \ , \\
u^{(1)}_k &= \rho_k \sin\theta_{k}^{(0)}\cos\theta_{k}^{(1)} \ ,\\
u^{(2)}_k &= \rho_k \sin\theta_{k}^{(0)}\sin\theta_{k}^{(1)}\cos\theta_{k}^{(2)} \ ,\\
u^{(3)}_k &= \rho_k \sin\theta_{k}^{(0)}\sin\theta_{k}^{(1)}\sin\theta_{k}^{(2)} \ ,
\end{align}
\end{subequations}
and the differential volume
\begin{gather}
du^{(0)}_k du^{(1)}_k du^{(2)}_k du^{(3)}_k = \rho_k^3 d\rho \sin^2\theta_{k}^{(0)} d\theta_{k}^{(0)} \sin\theta_{k}^{(1)} d\theta_{k}^{(1)} d\theta_{k}^{(2)} \ .
\end{gather}

Note that the distribution $P_\omega (\rho_\omega , \theta_{\omega}^{(0)}, \theta_{\omega}^{(1)}, \theta_{\omega}^{(2)})$ is not factorized in terms of real-valued
cartesian coordinates
$\omega^{(l)}$, $l=\overline{0,3}$ of $\omega_{12} = \omega^{(0)}+\mathfrak{i}\omega^{(1)}+\mathfrak{j}\omega^{(2)}+\mathfrak{k}\omega^{(3)}$.
 However, the marginal distributions of these parameters are identical and coincide with the distribution \eqref{eq:P(y)off-diag_GUE},
of the off-diagonal elements of the hermitian operator $\gamma_{ij}^{\alpha\beta}$ of the GUE case with $\sigma_2$ substituted for $\sigma_4$
\begin{equation}
P_{\omega^{(l)}}(x) = \bar P_{4,\sigma_4}(x)
\label{eq:P(x)prime_GSE}
\end{equation}

The diagonal elements $\alpha=\beta$ of the hopping operator
$\gamma_{ij}^{\alpha\beta} = 2{\Re} \omega_{ij}^{\alpha\beta} = 2\omega_{ij}^{\alpha\beta{(0)}}$
also obey the distribution \eqref{eq:P(x)prime_GSE},
while for the off-diagonal elements $\alpha\ne\beta$ we have
the marginal distributions of $\gamma^{(l)}$ given by \eqref{eq:P(x_M)} with $M=8$ and $\sigma_1$ replaced by $\sigma_4$
\begin{equation}
P_{\gamma^{(l)}}(z) = \bar P_{8,\sigma_4}(z) \ .
\end{equation}

\subsubsection{Sums of matrix elements}

As for the sum $\omega_M = \sum_k \omega_k$ of $M$ independent variables $\omega_k$
analogously to the previous sections the marginal distributions of the cartesian components $\omega_M^{(l)}$, $l=\overline{0,3}$ of
$\omega_M =\omega_M^{(0)}+\mathfrak{i}\omega_M^{(1)}+\mathfrak{j}\omega_M^{(2)}+\mathfrak{k}\omega_M^{(3)}$ are identical
and given by \eqref{eq:P(x_M)} with the corresponding effective $M$ and $\sigma_1$ replaced by $\sigma_4$
\begin{multline}\label{eq:P(x)sum_GUE}
P_{\omega_M^{(l)}}(X) = \\
\prod_{k=1}^M \int_{-\infty}^\infty P_\omega(\omega_k)\delta(X-\sum_k \omega_k^{(l)})\prod_{l'=0}^3 d\omega_k^{(l')}  =
\bar P_{2M,\sigma_4}(X) \ ,
\end{multline}
which approach the gaussian distribution with the variance scaling as $\sigma^2 = 2 M \sigma_4^4 \sim M /8 N^{2}$.

%\bibliography{references}

\end{document}